\documentclass[aps,showkeys,superscriptaddress,preprint,tightenlines]{revtex4}
\usepackage{hyperref}
\usepackage{times}
\usepackage{amsmath}
\usepackage{graphicx}
\usepackage{epsfig}
\usepackage{dcolumn}
\usepackage{bm}
\usepackage{color}
\usepackage{longtable}
\usepackage{array}
\usepackage{multirow}

\newcommand{\gev}{\rm GeV}

\newcommand{\gevcs}{{\rm GeV}/c^2}
\newcommand{\mev}{\rm MeV}
\newcommand{\mevcs}{{\rm MeV}/c^2}

\newcommand{\ev}{\rm eV}
\newcommand{\pbv}{{\rm pb}^{-1}}
\newcommand{\fbv}{{\rm fb}^{-1}}

\newcommand{\xyz}{\rm XYZ}
\newcommand{\xx}{X(3872)}
\newcommand{\y}{Y(4260)}

\newcommand{\zc}{Z_c(3900)}
\newcommand{\zcp}{Z_c(4020)}

\newcommand{\BR}{{\cal B}}

\newcommand{\piz}{\pi^0}

\newcommand{\hc}{h_c}
\newcommand{\pphc}{\pi^+\pi^- h_c}
\newcommand{\etac}{\eta_c}
\newcommand{\etacp}{\eta_c(2S)}

\newcommand{\psp}{\psi(2S)}
\newcommand{\psip}{\psi(2S)}
\newcommand{\pspp}{\psi(3770)}
\newcommand{\jpsi}{J/\psi}

\newcommand{\ccj}{\chi_{cJ}}

\newcommand{\cco}{\chi_{c1}}

\newcommand{\EE}{e^+e^-}

\newcommand{\LL}{\ell^+\ell^-}

\newcommand{\pp}{\pi^+\pi^-}

\newcommand{\kk}{K^+K^-}

\newcommand{\ks}{K_{S}^{0}}

\newcommand{\ccb}{c\bar{c}}
\newcommand{\ddb}{D\bar{D}}

\newcommand{\ppjpsi}{\pi^+\pi^- J/\psi}

\newcommand{\ddpi}{D^0D^{*-}\pi^+ + c.c.}
\newcommand{\ddstb}{D^*\bar{D}+c.c.}
\newcommand{\ddstbn}{D^{*0}\bar{D}^0+c.c.}

\newcommand{\beq}{\begin{equation}}
\newcommand{\eeq}{\end{equation}}
\newcommand{\bitm}{\begin{itemize}}
\newcommand{\eitm}{\end{itemize}}

\def\Journal#1#2#3#4{{#1} {\bf #2}, #3 (#4)}

\def\PRL{Phys. Rev. Lett.}
\def\PRD{Phys. Rev. D}

\begin{document}

\title{Charmonium and Charmoniumlike States at the BESIII Experiment}
\author{Chang-Zheng Yuan}
 \email{yuancz@ihep.ac.cn}
 \affiliation{Institute of High Energy Physics, Chinese Academy of Sciences,
 Beijing 100049, China}
 \affiliation{University of Chinese Academy of Sciences, Beijing 100049, China}

\begin{abstract}

Charmonium is a bound state of a charmed quark and a charmed
antiquark, and a {\it charmoniumlike} state is a resonant
structure that contains a charmed quark and antiquark pair but has
properties that are incompatible with a conventional charmonium
state. While operating at center-of-mass energies from 2 to
4.9~GeV, the BESIII experiment can access a wide mass range of
charmonium and charmoniumlike states, and has contributed
significantly in this field. We review BESIII results involving
conventional charmonium states, including the first observation of
M1 transition $\psp\to \gamma\etacp$ and the discovery of the
$\psi_2(3823)$; and report on studies of charmoniumlike states,
including the discoveries of the $\zc$ and $\zcp$ tetraquark
candidates, the resolution of the fine structure of the $\y$, the
discovery of the new production process $\EE\to \gamma \xx$, and
the uncovering of strong evidence for the commonality among the
$\xx$, $\y$, and $\zc$ states. The prospects for further research
at BESIII and proposed future facilities are also presented.

\end{abstract}


\maketitle

\section{Introduction}\label{Sec:intro}

In the conventional quark model, mesons are comprised of a quark
and anti-quark pair, while baryons are comprised of three quarks.
A bound state of a charmed quark ($c$) and a charmed antiquark
($\bar{c}$) is named charmonium. The first charmonium state, the
$\jpsi$, was discovered at BNL~\cite{ting} and at
SLAC~\cite{richter} in 1974, and since then, all the charmonium
states below the open-charm threshold and a few vector charmonium
states above the open-charm threshold have been
established~\cite{pdg}; the measured spectrum of states agrees
well with theoretical calculations based on QCD-inspired potential
models~\cite{eichten,godfrey,barnes}.

In addition to the charmonium states, the conventional quark model
describes almost all of the other hadrons that have been observed
to date quite well, including baryons and other mesons~\cite{pdg}.
Since the very beginning of the quark model, exotic hadronic
states with configurations not limited to two or three quarks have
been the subject of numerous theoretical proposals and
experimental searches~\cite{Jaffe:2004ph,klempt}. These proposed
exotic hadrons include hadron-hadron molecules,
diquark-diantiquark tetraquark states, hadro-quarkonia,
quark-antiquark-gluon hybrids, multi-gluon glueballs, and
pentaquark baryons.

Many charmonium and charmoniumlike states were discovered at the
BaBar~\cite{babar} and Belle~\cite{belle} $B$-factories during the
first decade of this century~\cite{PBFB}. Whereas some of these
are good candidates for conventional charmonium states, there are
other states that have properties that do not match those of any
of the unassigned $\ccb$ states, which may indicate that exotic
states have already been observed~\cite{reviews}. These candidate
exotic meson states are collectively called the $\xyz$ particles,
to indicate their underlying nature is still unclear. Although
this is not fully accepted within the high energy physics
community, practitioners in the field use $Z_Q({\rm xxxx})$ to
denote a quarkoniumlike state with mass roughly ${\rm
xxxx}$~MeV/$c^2$ that contains a heavy quark pair $Q\bar{Q}$ and
with non-zero isospin; $Y({\rm xxxx})$ for a vector quarkoniumlike
state, and $X({\rm xxxx})$ for states with other quantum numbers.

Although the BaBar~\cite{babar} and Belle~\cite{belle} experiments
finished data taking in 2008 and 2010, respectively, the data are
still used for various physics analyses. In 2008, two new
experiments: BESIII~\cite{bes3}, a $\tau$-charm factory experiment
at the BEPCII $\EE$ collider; and LHCb~\cite{lhcb_detector}, a
$B$-factory experiment at the LHC $pp$ collider, started data
taking, and have been contributing to the study of charmonium and
charmoniumlike states ever since.

The BESIII experiment at the BEPCII double ring $\EE$ collier
observed its first collisions in the $\tau$-charm energy region in
July 2008. The BESIII detector~\cite{bes3} is a magnetic
spectrometer with an effective geometrical acceptance of 93\% of
$4\pi$ and state-of-the-art subdetectors for high precision
charged and neutral particle measurements. After a few years of
running at center-of-mass (c.m.) energies for its well-defined
physics programs~\cite{BESIII_YB}, i.e., at the $\jpsi$ and $\psp$
peaks in 2009 and the $\pspp$ peak in 2010 and 2011, the BESIII
experiment began to collect data for the study of the $\xyz$
particles, a program that was not described in the BESIII Yellow
Book~\cite{BESIII_YB}. The first data sample was collected at the
$\psi(4040)$ resonance in May 2011 with an integrated luminosity
of about $0.5~\fbv$. This sample was used to search for the
production of the $\xx$ and the excited $P$-wave charmonium
spin-triplet states via $\psi(4040)$ radiative transitions. The
size of the sample was limited by the brief, one-month running
time following the $\pspp$ data taking in the 2010--2011 run.

In summer 2012, the LINAC of the BEPCII was upgraded so that the
highest beam energy was increased from $2.1$ to $2.3~\gev$, which
made it possible to collect data at higher c.m. energies (up to
$4.6~\gev$). A data sample of $525~\pbv$ was collected at a c.m.
energy of $4.26~\gev$ from December 14, 2012 to January 14, 2013,
with which the $\zc$ charged charmoniumlike state was
discovered~\cite{zc3900}. This observation changed the data
collection plan for the 2012--2013 run and had considerable impact
on the subsequent running schedule of the experiment; more data
points between $4.13$ and $4.60~\gev$ dedicated to the $\xyz$
related analyses were recorded~\cite{lum_songwm}. The highest beam
energy was further increased from $2.3$ to $2.45~\gev$ in summer
2019, making it possible to collect data at even higher c.m.
energies (up to $4.9~\gev$).

The data samples used for the $\xyz$ study cover the energy range
between 4.0 and 4.7~GeV, with a typical integrated luminosity of
$500~\pbv$ at each energy point. These data were also used for
charmonium studies together with a 448 million $\psp$ event
sample. Data samples with a total $826~\pbv$ integrated luminosity
at 104 energy points between $3.8$ and $4.6~\gev$~\cite{lum_Rscan}
was also used for the $\xyz$ study.

In this article, we review studies of charmonium and
charmoniumlike states from the BESIII~\cite{bes3} experiment. We
first introduce the study of conventional charmonium states and
then the $\xyz$ states. Finally, we discuss prospects for future
studies with the BESIII experiment, and also point out possible
studies at next generation facilities.

\section{Conventional charmonium states}

The search for new charmonium states has always been a high
priority topic. With the data taken at c.m. energies above 4~GeV,
it is possible to search for states predicted by the potential
models that are still unobseved~\cite{eichten,godfrey,barnes}.
These states include the excited $P$-wave spin-triplet states
$\chi_{cJ}(2P)$ ($J=0$, $1$, $2$), the excited $P$-wave
spin-singlet state $\hc(2P)$, the $D$-wave spin-triplet states
$\psi_J(1D)$ ($J=2$, $3$; $J=1$ state, the $\psi(3770)$, was
observed many years ago~\cite{pdg}), and the $D$-wave spin-singlet
state $\eta_{c2}(1D)$.

The predicted mass of the $D$-wave charmonium states (excluding
the $\psi(3770)$, which is, in fact, a mixture of the $1\,^3D_1$
and $2\,^3S_1$ vector states) is in the $3.81\sim 3.85$~GeV/$c^2$
range predicted by several phenomenological
calculations~\cite{eichten,godfrey,barnes,3d2-mass}. Since the
mass of $\psi_2(1D)$ is above the $D\bar{D}$ threshold but below
the $D\bar{D}^*$ threshold, and $\psi_2(1D)\to D\bar{D}$ violates
parity, the $\psi_2(1D)$ is expected to be narrow and its dominant
decay mode is $\psi_2(1D)\to
\gamma\chi_{c1}$~\cite{3d2-mass,ratio}. The $\psi_2(1D)$ state,
also called the $\psi_2(3823)$, was discovered at
BESIII~\cite{BES3x3823} in this final state, and the $\psi_3(1D)$
state was observed by LHCb in its decay into $D\bar{D}$ final
state~\cite{lhcb-3d3}.

The spin-triplet charmonium states are produced copiously in $\EE$
annihilation and in $B$ decays and, thus, they are understood much
better than the spin-singlet charmonium states, including the
lowest lying $S$-wave state, the $\etac$, its radial excited
partner the $\etacp$, and the $P$-wave spin-singlet state the
$\hc$. Since these three states are all produced in $\psp$ decays,
the world largest $\psp$ data sample at BESIII made it possible to
study their properties with improved precision. In addition, the
unexpected large production cross section for $\EE\to \pp\hc$ in
BESIII energy region~\cite{bes3_pipihc_lineshape} opened a new
mechanism of studying the $\hc$ and $\etac$ (from $\hc\to \gamma
\etac$), and BESIII contributed world best measurements of the
properties of these states~\cite{bes3_etacbr}. We report here the
observation of the M1 transition $\psp\to \gamma \etacp$ at
BESIII~\cite{bes3_etacp}, a transition that has been sought for
since the first generation BES experiment in 1980's.

\subsection{\boldmath Discovery of M1 transition $\psp\to \gamma\etacp$}

The production of the $\etacp$ through a radiative transition from
the $\psp$ involves a charmed-quark spin-flip and, thus, proceeds
via a magnetic dipole (M1) transition. The branching fraction has
been calculated by many authors, with predictions in the range
$\BR(\psp\to \gamma \etacp)= (0.1-6.2)\times
10^{-4}$~\cite{gaoky,zhaoq,mabq}. Experimentally, this transition
has been searched for by Crystal Ball~\cite{cbal},
BES~\cite{ycz,initial_psp}, and CLEO~\cite{etacp_cleo_c}. No
convincing signal was observed by any of these experiments.

\begin{figure*}[htbp]
\centering
  \includegraphics[width=0.49\textwidth]{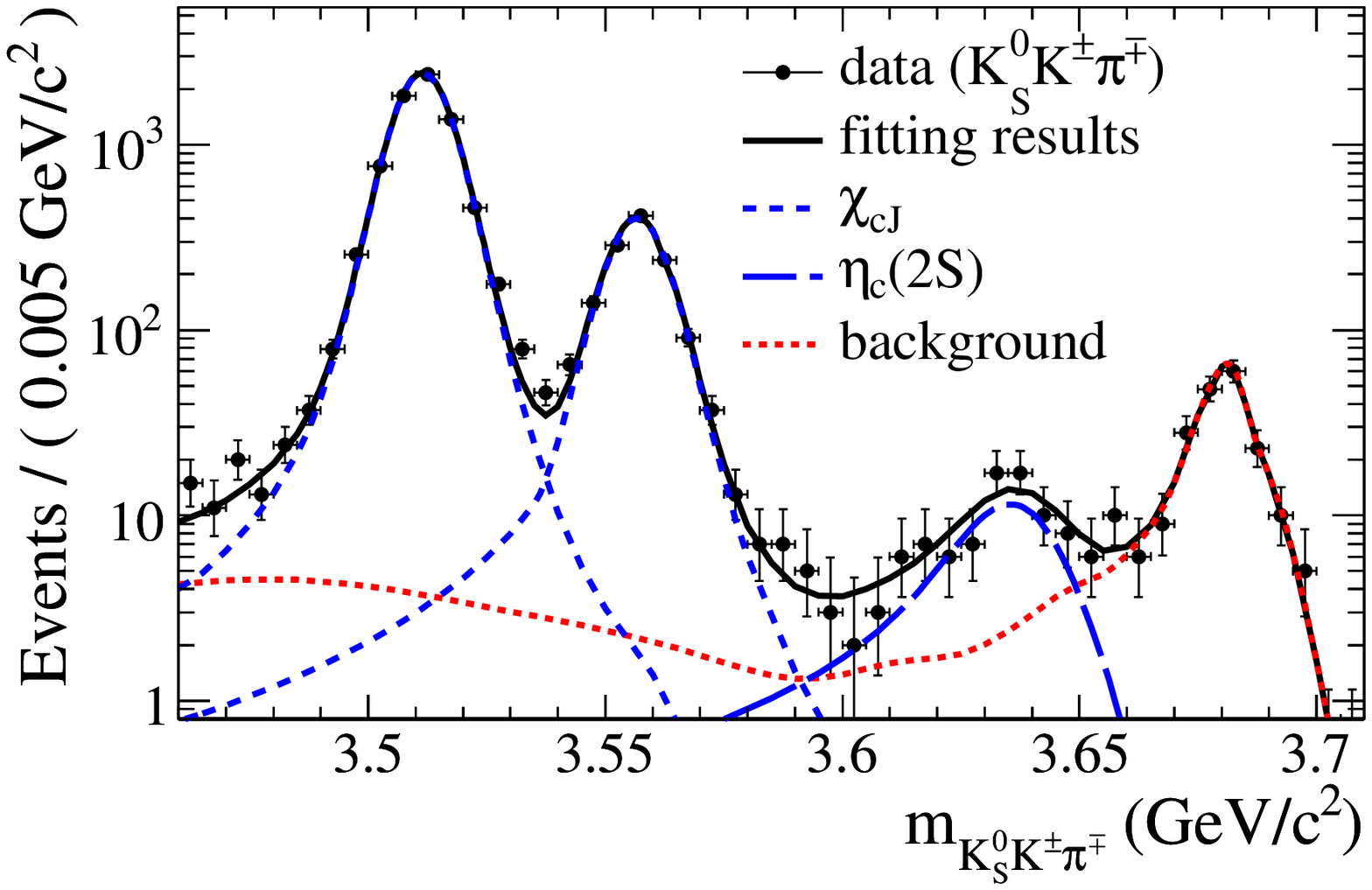}
  \includegraphics[width=0.49\textwidth]{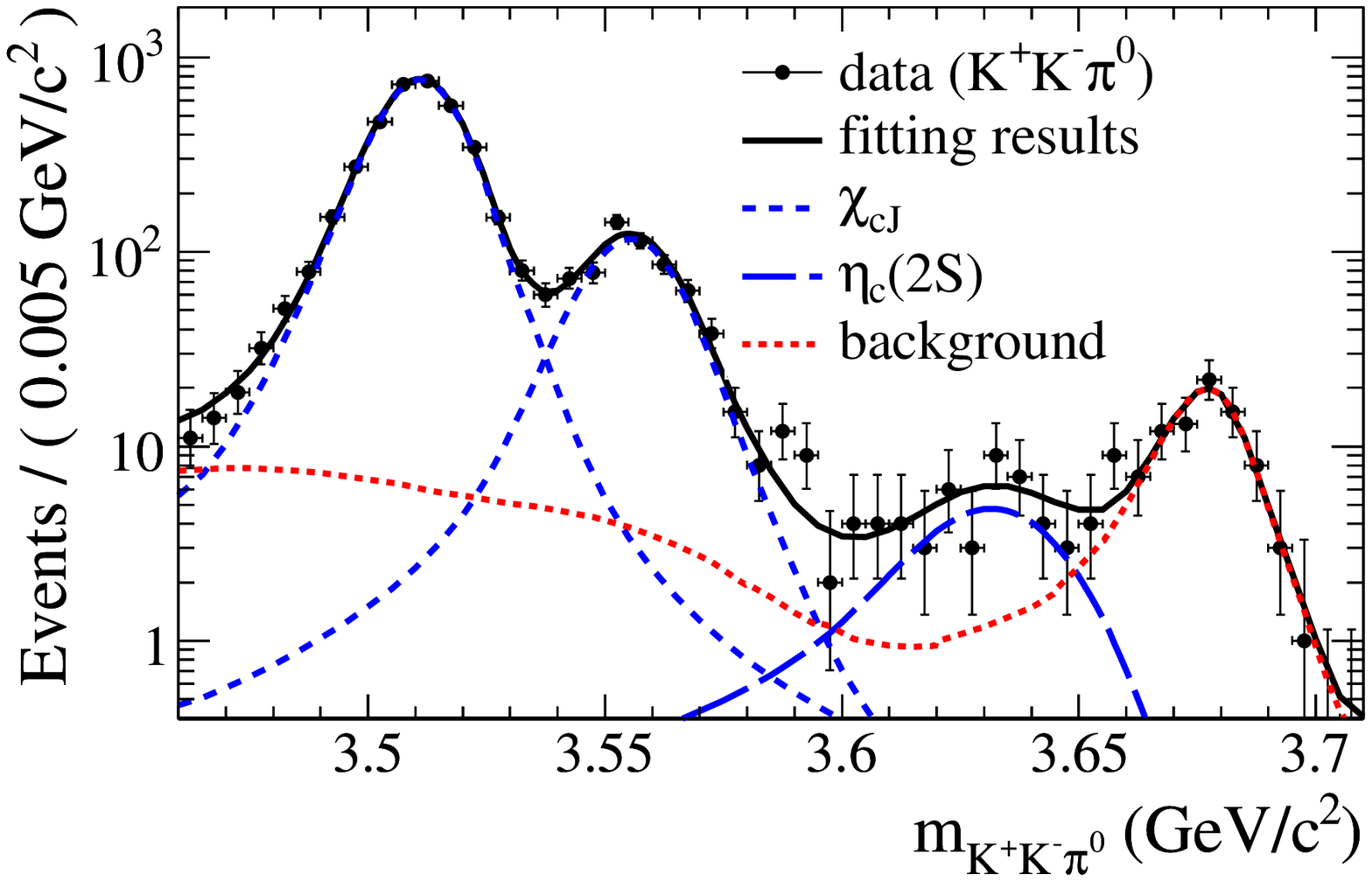}
\caption{The fit to the invariant-mass spectra for $\ks
K^\pm\pi^\mp$ (left panel) and $\kk\piz$ (right panel). Dots with
error bars are data, and the curves are total fit and each
component. The lowest peaks correspond to the $\etacp$ signals.}
\label{pic_fit_etacp}
\end{figure*}

With a sample of 106 million $\psp$ events collected at BESIII,
the process $\psp\to \gamma\etacp$ was observed for the first time
with $\etacp\to \ks K^\pm\pi^\mp$ and $\kk\piz$ modes. The final
$K\bar{K}\pi$ mass spectra and the fit results are shown in
Fig.~\ref{pic_fit_etacp}. The fit yields for the number of the
$\etacp$ signal events are $81\pm 14$ for the $\ks K^\pm\pi^\mp$
mode and $46\pm 11$ for the $\kk\piz$ mode; the overall
statistical significance of the signal is larger than
$10\sigma$~\cite{bes3_etacp}.

The mass of the $\etacp$ is measured to be $(3637.6\pm 2.9\pm
1.6)$~MeV/$c^2$, the width $(16.9\pm 6.4\pm 4.8)$~MeV, in good
agreement with the PDG world average values~\cite{pdg}, and the
product branching fractions $\BR(\psp\to \gamma\etacp)\times
\BR(\etacp\to K\bar K\pi) = (1.30\pm 0.20\pm 0.30)\times 10^{-5}$.
Combining this result with a BaBar measurement of $\BR(\etacp\to
K\bar K \pi)$, the M1 transition rate is determined to be
$\BR(\psp\to\gamma\etacp) = (6.8\pm 1.1_\mathrm{stat}\pm
4.5_\mathrm{sys})\times 10^{-4}$. This agrees with theoretical
calculations~\cite{gaoky,zhaoq,mabq} and naive estimates based on
the $\jpsi\to \gamma \eta_c$ transition~\cite{etacp_cleo_c}.

This study benefited from the BESIII detector's high resolution
electromagnetic calorimeter, which makes the detection of the
radiative photon with 50~MeV energy possible~\cite{bes3}. Given
the tiny transition rate and the low photon energy, it is
understandable why this transition was not observed in previous
studies~\cite{cbal,ycz,etacp_cleo_c}. This is the third M1
transition observed in a charmonium system (the other two are
$\jpsi\to \gamma\etac$ and $\psp\to \gamma\etac$ observed in
1980~\cite{cbal_etac}); improved measurements of these transitions
and discovery of more M1 transitions would improve the
understanding of the high order effects involved in these
transitions~\cite{barnes}.

\subsection{\boldmath Observation of $\psi_2(3823)$} \label{Sec:x3823}

\begin{figure}[htbp]
\centering
\includegraphics[height=5.0cm]{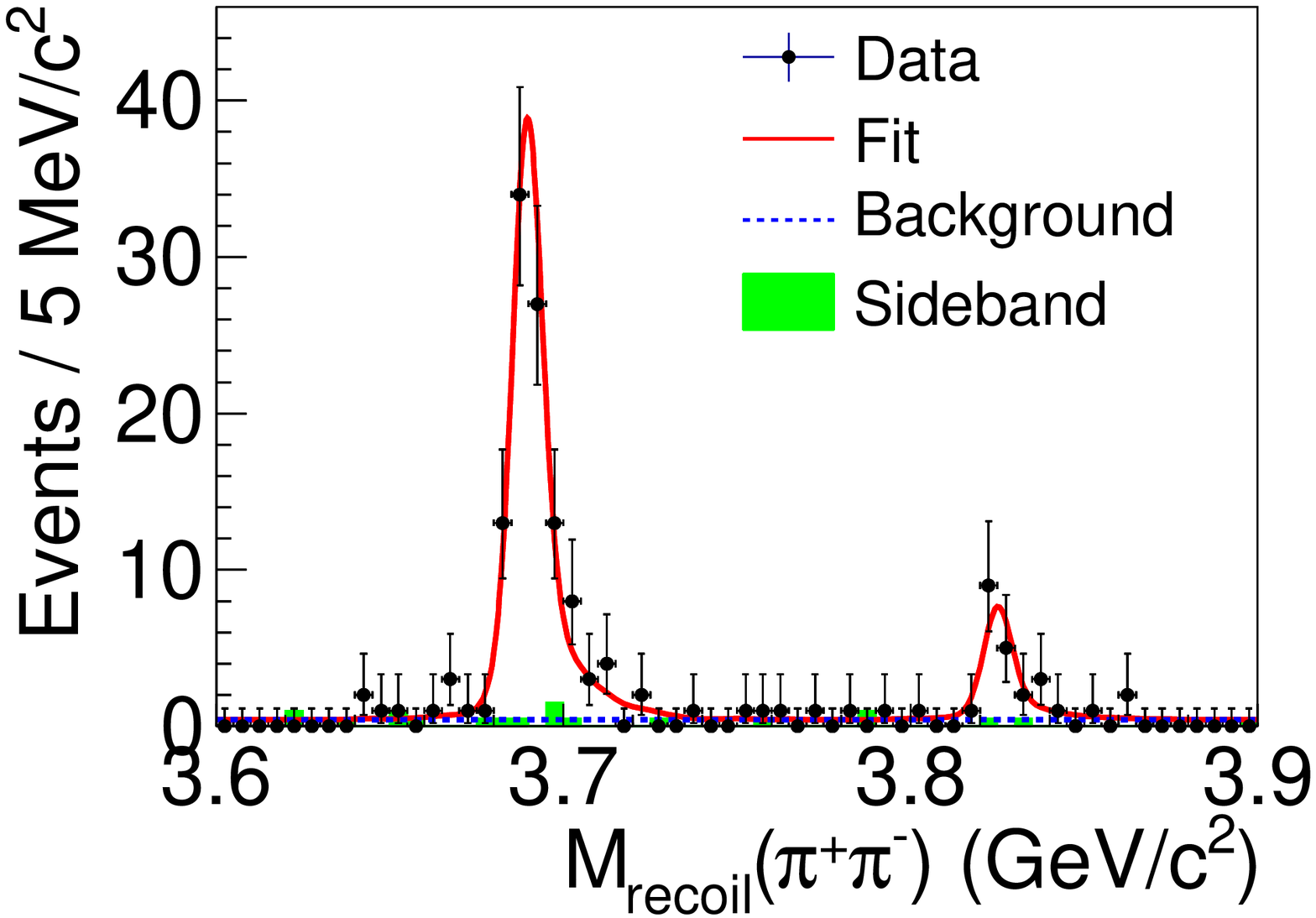}
\includegraphics[height=5.0cm]{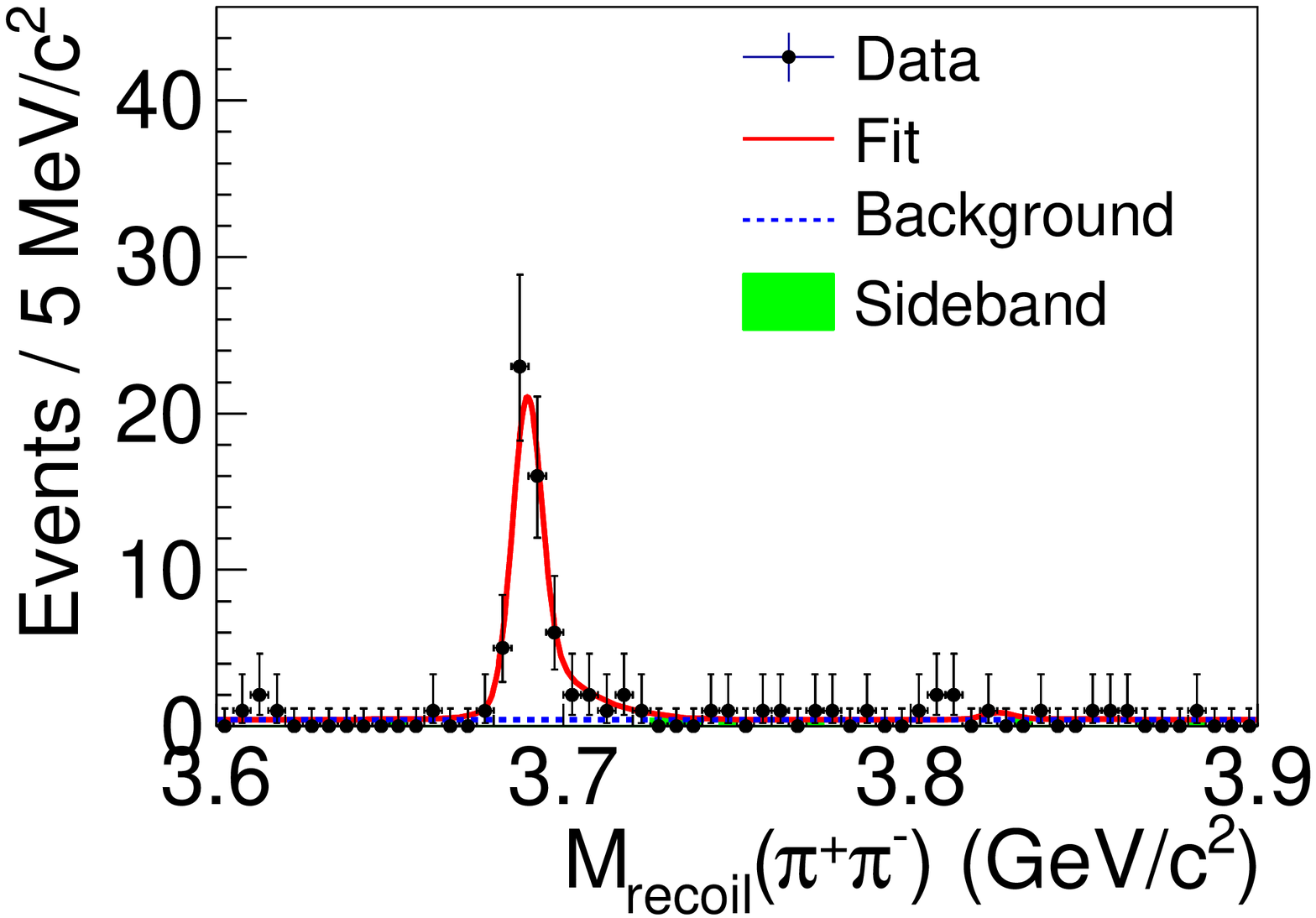}
\caption{Simultaneous fit to the $M_{\rm recoil}(\pp)$
distribution of $\gamma\chi_{c1}$ events (left) and
$\gamma\chi_{c2}$ events (right), respectively~\cite{BES3x3823}.
The small peak in the left panel is the $\psi_2(3823)$ signal.
Dots with error bars are data, red solid curves are total fit,
dashed blue curves are background, and the green shaded histograms
are $\jpsi$ mass sideband events.} \label{X-fit}
\end{figure}

The processes of $\EE\to \pp \gamma\chi_{c1,2}$ are studied at
BESIII experiment using $4.1~\fbv$ data samples collected at c.m.
energies from $4.23$ to $4.60~\gev$~\cite{BES3x3823}. The
$\chi_{c1,2}$ are reconstructed via their decays into
$\gamma\jpsi$, with $\jpsi$ to $\LL$ ($\ell=e, \mu$). A clear
signal is observed as a $19\pm 5$ event peak in the
$\gamma\chi_{c1}$ invariant mass distribution that is evident in
Fig.~\ref{X-fit}(1eft); its mass is determined to be $(3821.7\pm
1.3\pm 0.7)~\mevcs$, and its properties are in good agreement with
the $\psi(1^3D_2)$ charmonium state. The statistical significance
of the $\psi_2(3823)$ signal is $6.2\sigma$. The upper limit on
the intrinsic width of the $\psi_2(3823)$ is determined as
$16~\mev$ at the 90\% confidence level (C.L.). This observation is
in good agreement with the $3.8\sigma$ ``evidence'' in $B$ decays
reported by Belle~\cite{belle-3d2}. For the $\gamma\chi_{c2}$
mode, no significant $\psi_2(3823)$ signal is observed
(Fig.~\ref{X-fit}(right)), and an upper limit on its production
rate is determined. BESIII obtains the ratio
$\frac{\BR[\psi_2(3823)\to \gamma\chi_{c2}]} {\BR[\psi_2(3823)\to
\gamma\chi_{c1}]}<0.42$ at the 90\% C.L., which also agrees with
expectations for the $\psi(1\,^3D_2)$ state~\cite{ratio}.

With the observation of three $D$-wave spin-triplet states, their
center-of-gravity, 3822~MeV/$c^2$, is a good estimation of the
mass of the $D$-wave spin-singlet state, $\eta_{c2}(1D)$. Since it
cannot decay into open charm final states, the $\eta_{c2}(1D)$ is
expected to be very narrow, and the identification of it should be
clear, if it is produced with large enough rate, in $\EE\to \gamma
\eta_{c2}(1D)$ or $\EE\to \pp h_c(2P)\to \pp\gamma\eta_{c2}(1D)$.

\section{Exotic charmoniumlike states}

A revival of the study of charmonium spectroscopy occurred in the
early 21st century when the BaBar and Belle $B$-factories started
accumulating large data samples at $\Upsilon(4S)$ peak. The high
luminosity at these $B$-factories enabled studies of charmonium
states that are produced in a variety of ways, including $B$
decays, initial-state-radiation (ISR) processes, double-charmonium
production, two-photon processes, etc. While the discovery of the
conventional charmonium states such as $\etacp$ and
$\chi_{c2}(2P)$ were more-or-less rountine, the observations of
the $\xx$ by Belle in 2003~\cite{bellex} and the $\y$ by BaBar in
2005~\cite{babar_y4260}, the first of the $\xyz$ mesons, came as
big surprises; although these new states decay to final states
that contain both a $c$- and a $\bar{c}$-quark, they have
properties that do not match those of any $\ccb$
meson~\cite{reviews}.

All studies of $\xyz$ states at the $B$-factories suffer from low
statistics and limited precision. In contrast, BESIII can tune the
c.m. energy to match the peaks of the $Y$ states, where event
rates are high enough to facilitate precise measurements of their
resonance parameters and search for new states among their decay
products.

\subsection{\boldmath New insights into the $Y$ states}\label{Sec:y}

The $Y$ states, such as the $Y(4260)$~\cite{babar_y4260}, the
$Y(4360)$~\cite{babar_y4360,belle_y4660} and the
$Y(4660)$~\cite{belle_y4660}, are produced directly or via the ISR
process in $\EE$ annihilation and, thus, are vectors with quantum
numbers $J^{PC}=1^{--}$. These states have strong couplings to
hidden-charm final states in contrast to the established vector
charmonium states in the same mass region, such as the
$\psi(4040)$, $\psi(4160)$, and $\psi(4415)$, which dominantly
couple to open-charm meson pairs~\cite{bes2_psis,uglov-kmatrix}.

In potential models, five vector charmonium states are expected to
be in the mass region between 4.0 and 4.7~GeV/$c^2$, namely the
$\psi(3S)$, $\psi(2D)$, $\psi(4S)$, $\psi(3D)$, and
$\psi(5S)$~\cite{3d2-mass}, with the first three identified with
the well-established $\psi(4040)$, $\psi(4160)$, and $\psi(4415)$
charmonium mesons; the masses of the as yet undiscovered
$\psi(3D)$ and $\psi(5S)$ are expected to be higher than
4.4~GeV/$c^2$. However, six vector states in this mass region have
been identified, as listed above. These makes the $Y(4260)$,
$Y(4360)$ and perhaps the $Y(4660)$ states good candidates for new
types of exotic particles and this stimulated many theoretical
interpretations, including tetraquark states, molecular states,
hybrid states, or hadro-charmonia~\cite{reviews}.

The $\y$ was first observed at the $B$-factories as a distinct
peak in the $\pp\jpsi$ invariant mass distribution for
ISR-produced $\EE\to \gamma_{\rm ISR}\pp\jpsi$
events~\cite{belley,babar_y4260}. Improved measurements from both
BaBar~\cite{babar_y4260_new} and Belle~\cite{belley_new} with
their full data samples confirmed the existence of both the $\y$
and a non-$\y$ component in $\EE\to\pp\jpsi$ around $4.0~\gev$,
but the line-shape was parameterized with different models. The
parameters of the $\y$ determined by fit to the combined data from
the two $B$-factory experiments and the CLEO
measurements~\cite{cleo_isry} are $M_{\y}=(4251\pm 9)~\mevcs$ and
$\Gamma_{\y}=(120\pm 12)~\mev$~\cite{pdg2016}. High precision
BESIII measurements of the direct cross section for the $\y$
production in different final states supply new insight into its
nature. These measurements include: $\EE\to
\pp\jpsi$~\cite{bes3_pipijpsi_lineshape}, $\EE\to
\pp\hc$~\cite{bes3_pipihc_lineshape}, $\EE\to
\omega\ccj$~\cite{bes3_omegachic0,bes3_omegachic0_new}, $\EE\to
\ddpi$~\cite{bes3_ddstarpi}, and so on~\cite{bes3_white}.

\begin{figure*}[htbp]
\centering
\includegraphics[width=0.8\textwidth]{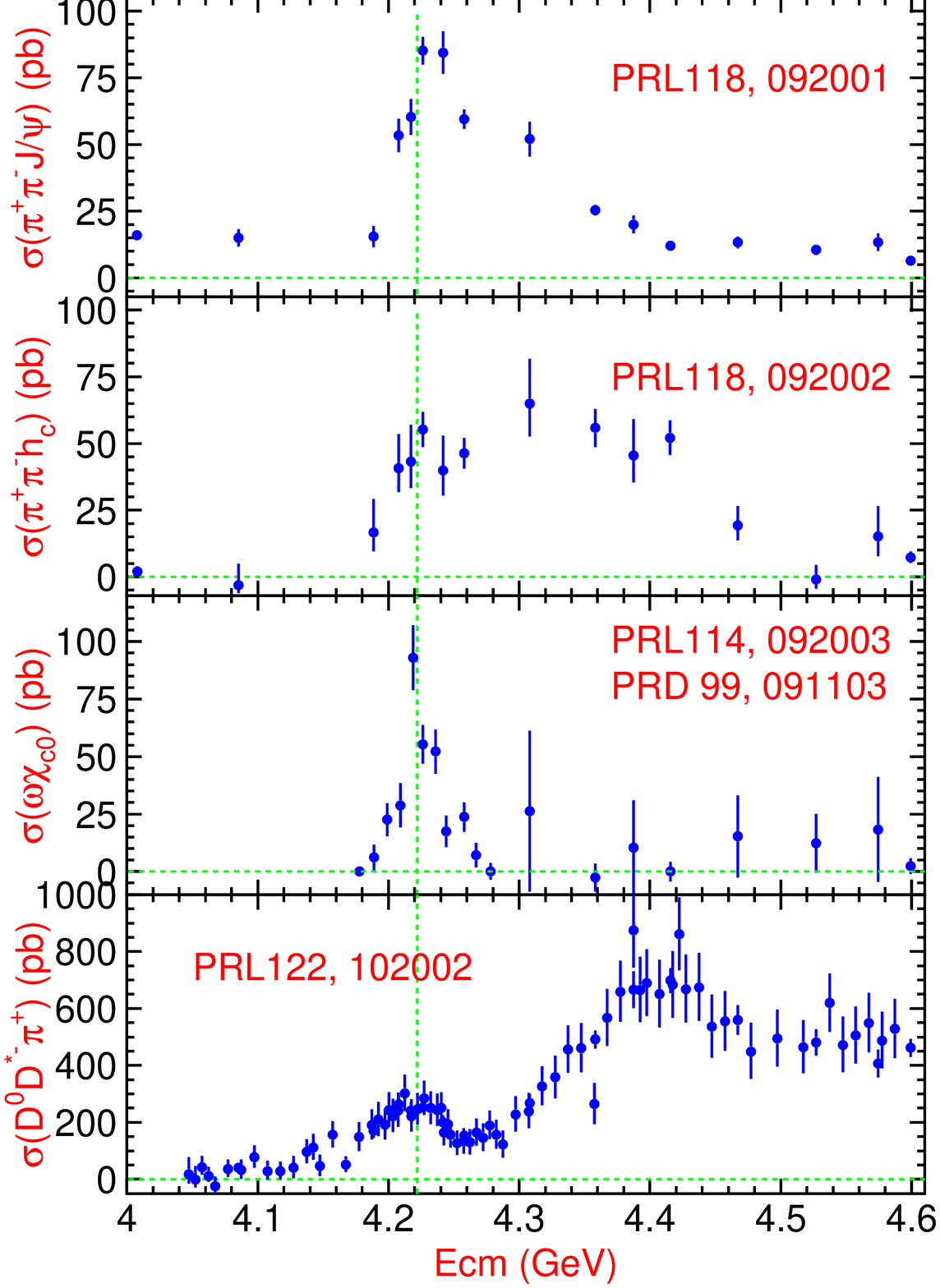}
\caption{From top to bottom are the measured cross sections of
$\EE\to \ppjpsi$~\cite{bes3_pipijpsi_lineshape}, $\EE\to
\pp\hc$~\cite{bes3_pipihc_lineshape}, $\EE \to \omega
\chi_{c0}$~\cite{bes3_omegachic0,bes3_omegachic0_new}, and $\EE\to
\ddpi$~\cite{bes3_ddstarpi}. Dots with error bars are the data and
the dotted vertical line is the peak of the $Y(4220)$.}
\label{xsec-fit}
\end{figure*}

Figure~\ref{xsec-fit} shows the measured cross sections for each
of these final states. The $\y$ structure is evident, but its line
shape is, in fact not well described by a single Breit-Wigner (BW)
resonance function. Instead, its line-shape is peaked at around
$4.22~\gev$, which is substantially lower than the average value
from previous measurements~\cite{pdg2016}, and a distinct shoulder
is observed on its high-mass side that is especially pronounced in
the $\pp\jpsi$ mode. In order to describe this line shape, two
resonant structures in the $\y$ peak region are needed. The lower
one has a mass of $(4222.0\pm 3.1\pm 1.4)~\mevcs$ and a width of
$(44.1\pm 4.3\pm 2.0)~\mev$, while the higher one has a mass of
$(4320.0\pm 10.4\pm 7.0)~\mevcs$ and a width of
$(101.4^{+25.3}_{-19.7}\pm 10.2)~\mev$. The mass of the first
resonance is $\sim 30~\mevcs$ lower than the world average value
at that time~\cite{pdg2016} for the $\y$ and its width is about a
factor of three narrower. The second resonance is observed in the
$\EE\to \ppjpsi$ process for the first time. The resonance
parameters for the $\y$ structures are also measured in other
decay channels and listed in Table~\ref{tab:Y4260}.

\begin{table}[htbp]
\caption{Resonance parameters of the $Y(4220)$ from different
modes measured at BESIII. The cross sections measured at c.m.
energy of 4.226~GeV are also listed.}
    \label{tab:Y4260}
    \centering
    \begin{tabular}{cccc}
    \hline\hline
           Mode              &  Mass (GeV/$c^2$)              & Width (MeV)    & $\sigma$ at $\sqrt{s}=4.226$~GeV (pb) \\\hline
    $\EE\to \pp\jpsi$        & $4222.0\pm 3.1\pm 1.4$         & $44.1\pm 4.3\pm 2.0$         & $85.1\pm 1.5\pm 4.9$  \\
    $\EE\to \pp\hc$          & $4218.4^{+5.5}_{-4.5}\pm 0.9$  & $66.0^{+12.3}_{-8.3}\pm 0.4$ & $55.2\pm 2.6\pm 8.9$  \\
    $\EE\to \omega\chi_{c0}$ & $4218.5\pm 1.6\pm 4.0$         & $28.2\pm 3.9\pm 1.6$         & $55.4\pm 6.0\pm 5.9$  \\
    $\EE\to \pi^+ D^0 D^{*-}+c.c.$  & $4228.6\pm 4.1\pm 6.3$  & $77.0\pm 6.8\pm 6.3$         & $252\pm 5\pm 15$  \\
    \hline\hline
    \end{tabular}
\end{table}

Since the resonant structure around $4.2~\gevcs$ is present in all
of the above channels with similar resonance parameters, the
authors of Ref.~\cite{gaoxy} applied a combined fit to the
measured cross sections to determine the resonance parameters of
the low-mass $Y(4220)$ peak with a resultant mass of $(4219.6 \pm
3.3 \pm 5.1)~\mevcs$ and width of $(56.0 \pm 3.6 \pm 6.9)~\mev$.
These values are very different from those obtained from previous
experiments~\cite{pdg2016}. The fit also gives the product of the
leptonic decay width and the decay branching fraction to the
considered final state. After accounting for the unmeasured
isospin partners of the measured channels, a lower limit on the
leptonic partial width of the $Y(4220)$ is determined to be
$\Gamma_{\EE} > (29.1\pm 7.4)~\ev$, where the error is the
combined fit error and those from different fit scenarios. The
authors of Ref.~\cite{zhenghq} analyzed BESIII, Belle, and BaBar
data on charmonium as well as open charm final states, and a
leptonic width of $O(0.1\sim 1)$~keV is obtained. This partial
width is much larger than LQCD predictions for a hybrid vector
charmonium state~\cite{chenying}.

In spite of the limited experimental information that had been
available between the time of its discovery in 2005 and the recent
BESIII measurements, the $\y$ has attracted considerable
attention. The BESIII measurements of its production, decay, and
line shape in a variety of final states enable more sophisticated
theoretical investigations and some analyses have been performed,
such as those in Ref.~\cite{zhenghq} and those quoted in
Ref.~\cite{reviews}. The presence of the nearby
$D_s^{*+}D_s^{*-}$, $D\bar{D}_1(2420)$, and $\omega\chi_{cJ}$
production thresholds, and its mass overlap with the $\psi(4160)$
and $\psi(4415)$ conventional charmonium states complicate its
interpretation. Joint experimental and theoretical efforts will
likely be required to gain a full understanding of the nature of
this state.

\subsection{\boldmath Discovery of the iso-triplet charmoniumlike $\zc$ and $\zcp$ states}
\label{Sec:z}

Searching for charged charmoniumlike states is one of the most
promising ways of establishing the existence of the exotic
hadrons, since such a state must contain at least four quarks and,
thus, could not be a conventional meson. These searches have been
concentrated on decay final states that contain one charged pion
and a charmonium state, such as the $\jpsi$, $\psp$, and $\hc$,
since they are narrow and their experimental identification is
relatively unambiguous.

The first reported charged charmoniumlike state, the
$Z_c(4430)^-$, was found in the $\pi^-\psp$ invariant mass
distribution in $B\to K\pi^-\psp$ decays by the Belle experiment
in 2008~\cite{Belle_zc4430,Belle_zc4430pwa}. It was confirmed by
the LHCb experiment seven years later~\cite{LHCb_zc4430}. The
$\zc^-$ was observed in the $\pi^-\jpsi$ invariant mass
distribution in the study of $\EE\to \ppjpsi$ at
BESIII~\cite{zc3900} and Belle~\cite{belley_new} experiments, and
the $\zcp^-$ was observed in the $\pi^-\hc$ system in $\EE\to
\pphc$~\cite{zc4020} only at BESIII.

\subsubsection{Observation of the $\zc$}
\label{Sec:zc-pijpsi}

\begin{figure}[htbp]
\begin{center}
\includegraphics[height=5.0cm]{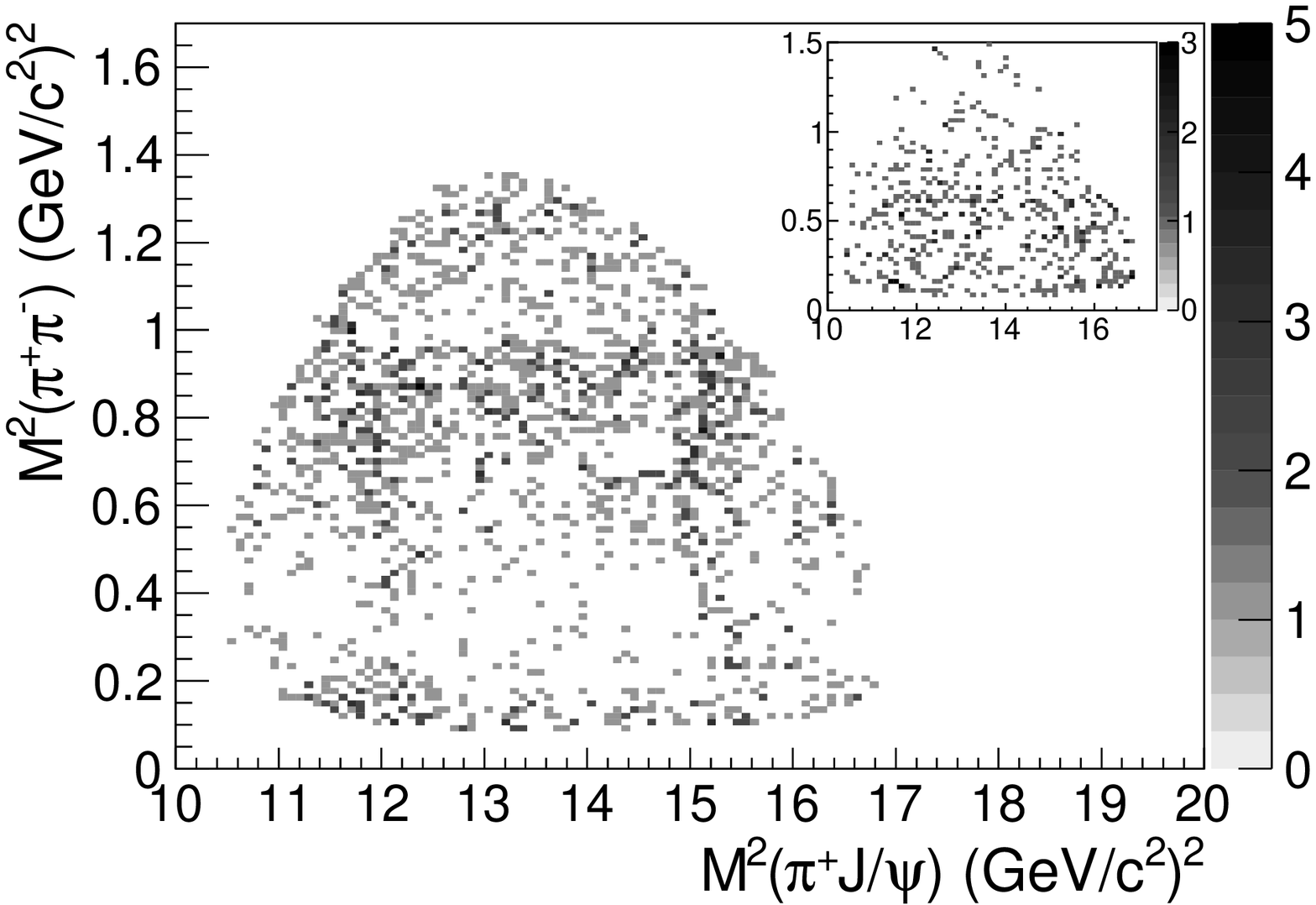}
\includegraphics[height=5.0cm]{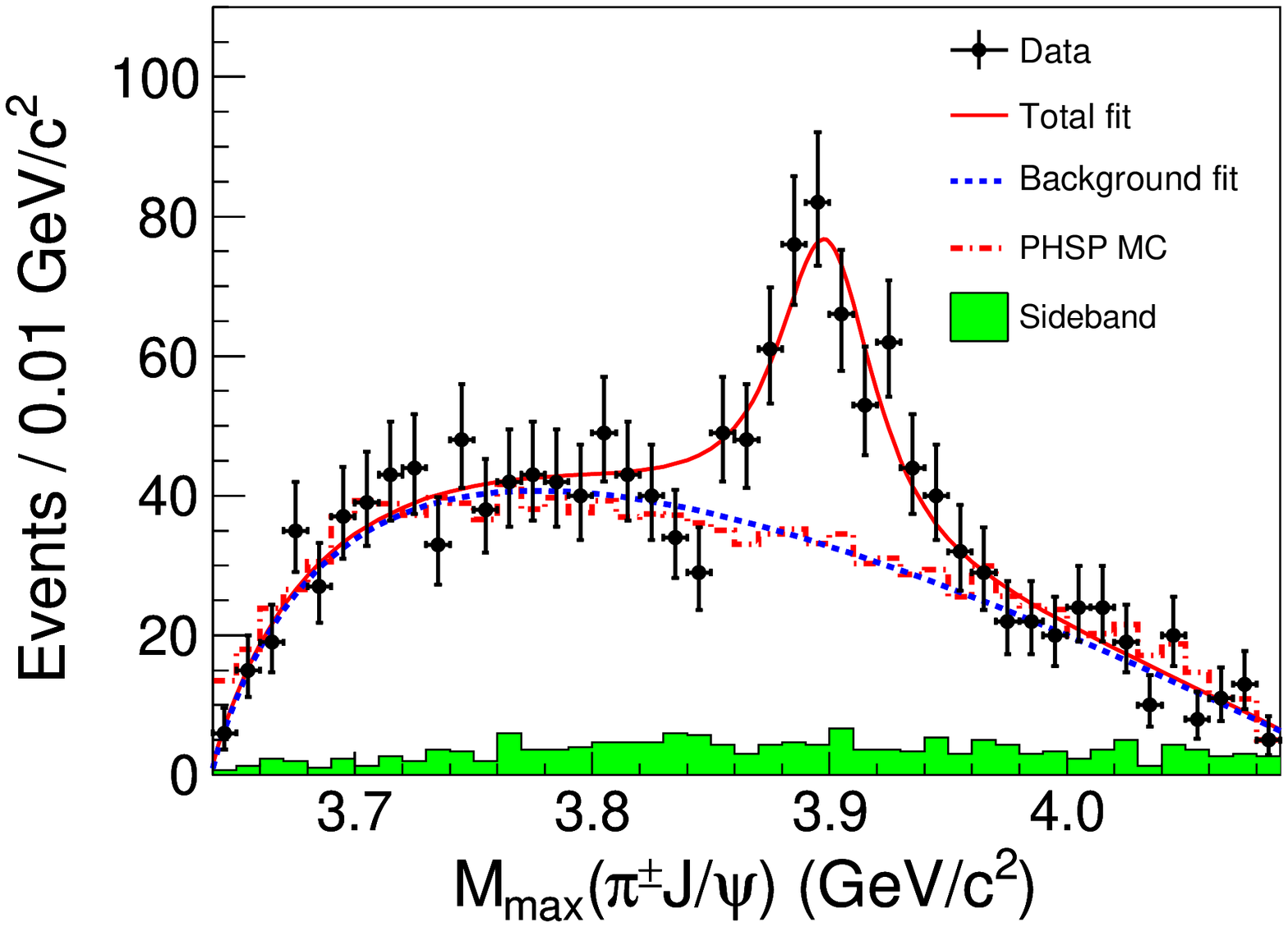}
\caption{Dalitz plot for selected $\EE\to \ppjpsi$ events in the
$\jpsi$ signal region (left, the inset show background events from
the $\jpsi$ mass sidebands) and the $\zc$ signal in the
$M_{\mathrm{max}}(\pi J/\psi)$ (right). Points with error bars are
data, the curves are the best fit, the dashed histograms are the
phase space distributions and the shaded histograms are the
non-$\ppjpsi$ background estimated from the normalized $\jpsi$
sidebands.} \label{zc}
\end{center}
\end{figure}

The BESIII experiment studied the $\EE\to \ppjpsi$ process using a
$525~\pbv$ data sample at a c.m. energy of
$4.26~\gev$~\cite{zc3900}. About $1500$ signal events were
observed and the cross section was measured to be $(62.9\pm 1.9
\pm 3.7)$~pb, which agrees with the previous existing results from
the Belle~\cite{belley} and BaBar~\cite{babar_y4260_new}
experiments. The intermediate states in this three-body system
were studied by examining the Dalitz plot of the selected
candidate events, as shown in Fig.~\ref{zc}.

In addition to the known $f_0(500)$ and $f_0(980)$ structures in
the $\pp$ system, a structure at around $3.9~\gevcs$ was observed
in the $\pi^\pm \jpsi$ invariant mass distribution with a
statistical significance larger than $8\sigma$, which is referred
to as the $\zc$. A fit to the $\pi^\pm\jpsi$ invariant mass
spectrum (see Fig.~\ref{zc}) determined its mass to be $(3899.0\pm
3.6\pm 4.9)~\mevcs$ and its width to be $(46\pm 10\pm 20)~\mev$.

An article from the Belle experiment that was released subsequent
to the BESIII paper reported the observation of the $\zc$ state
(referred to as $Z(3900)^+$ in the Belle paper) produced via the
ISR process with a mass of $(3894.5\pm 6.6\pm 4.5)~\mevcs$ and a
width of $(63\pm 24\pm 26)~\mev$ with a statistical significance
larger than $5.2\sigma$~\cite{belley_new}. These observations were
later confirmed by an analysis of CLEO-c data at a c.m. energy of
$4.17~\gev$~\cite{seth_zc}, with a mass and width that agree with
the BESIII and Belle measurements.

BESIII studied the spin-parity of the $\zc$ with a partial wave
analysis (PWA) of about 6000 $\EE\to \pp\jpsi$ events at
$\sqrt{s}=4.23$ and $4.26~\gev$~\cite{zc3900_jpc}. The fit
indicated that the spin-parity $J^P=1^+$ assignment for the $\zc$
is favored over other quantum numbers ($0^-$, $1^-$, $2^-$, and
$2^+$) by more than 7$\sigma$.

The $\zc$ mass determined from its $\pi\jpsi$ invariant mass
distribution is slightly above the $D\bar{D}^*$ mass threshold.
The open-charm decay $\zc^\pm\to (D\bar{D}^*)^\pm$ was observed
with much larger rate than that to
$\pi\jpsi$~\cite{zc3885st,zc3885dt}, and the pole mass and width
were determined with high precision to be $(3882.2\pm1.1\pm
1.5)~\mevcs$ and $(26.5\pm 1.7\pm 2.1)~\mev$, respectively.

\begin{figure}[htbp]
  \centering
     \includegraphics[width=0.6\textwidth]{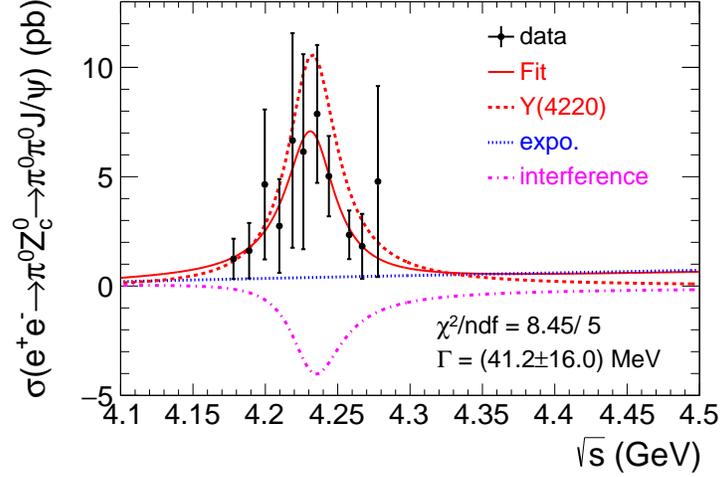}
\caption{The cross sections of $\EE\to\piz\zc^0\to
\piz\piz\jpsi$~\cite{zc0-lineshape}. Points with error bars are
data, the red solid curve is the total fit result, the red-dashed
(blue-dotted) curve is the resonant (non-resonant) component, and
the magenta dash-dotted line represents the interference of the
two components.} \label{fig:zc0}
\end{figure}

In both the QCD tetraquark and the molecular pictures, the
$\zc^{\pm}$ states are the $I_3=\pm 1$ members of an isospin
triplet. BESIII confirmed this by observing their neutral, isospin
$I_3 = 0$ partners: the $\zc^0$, in both the
$\piz\jpsi$~\cite{Ablikim:2015tbp} and
$(D\bar{D}^*)^0$~\cite{Ablikim:2015gda} decay modes. These
observations establish the $\zc$ as isovector states with even
$G$-parity. From a PWA to the $\EE\to \piz\piz\jpsi$ data in the
vicinity of the $\y$ resonance, it is found that the cross section
line shape of $\EE\to \piz\zc^0\to \piz\piz\jpsi$ is in agreement
with that of the $Y(4220)$ (see
Fig.~\ref{fig:zc0})~\cite{zc0-lineshape}.

BESIII also searched for the $\zc$ isospin violating decay mode
$\eta\jpsi$~\cite{Zc_jpsieta_bes3} as well as to the light hadron
final states $\omega\pi$~\cite{Zc_omegapi_bes3}, $K\bar{K}\pi$ and
$K\bar{K}\eta$~\cite{Zc_kkpi_bes3}, these were not observed and
the upper limits of the decay rates are one order of magnitude or
even smaller than that for $\zc\to \pi\jpsi$, as naively expected.

\subsubsection{Observation of the $\zcp$}
\label{Sec:zcp-pihc}

\begin{figure}[htbp]
\begin{center}
\includegraphics[width=0.45\textwidth]{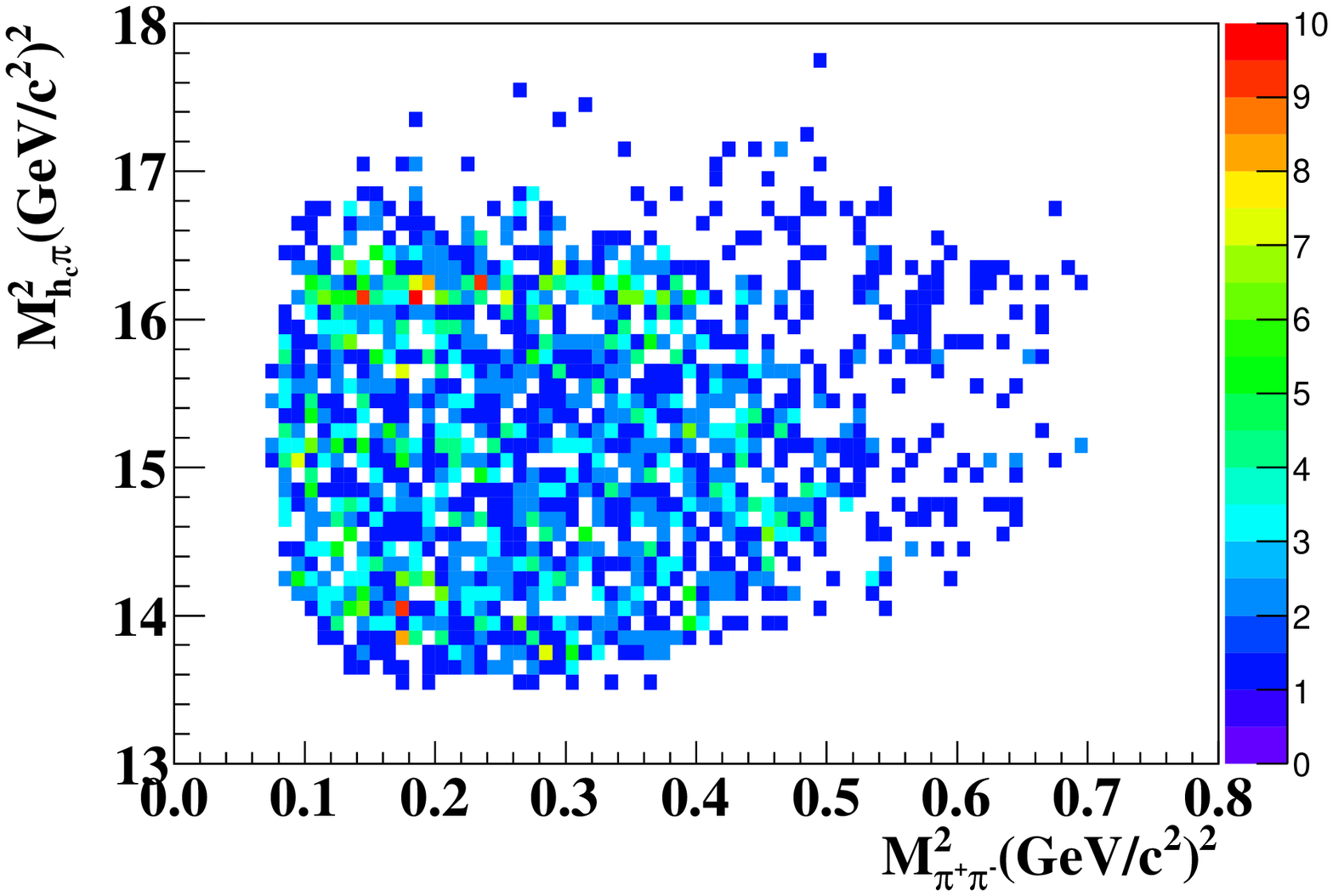}
\includegraphics[width=0.45\textwidth]{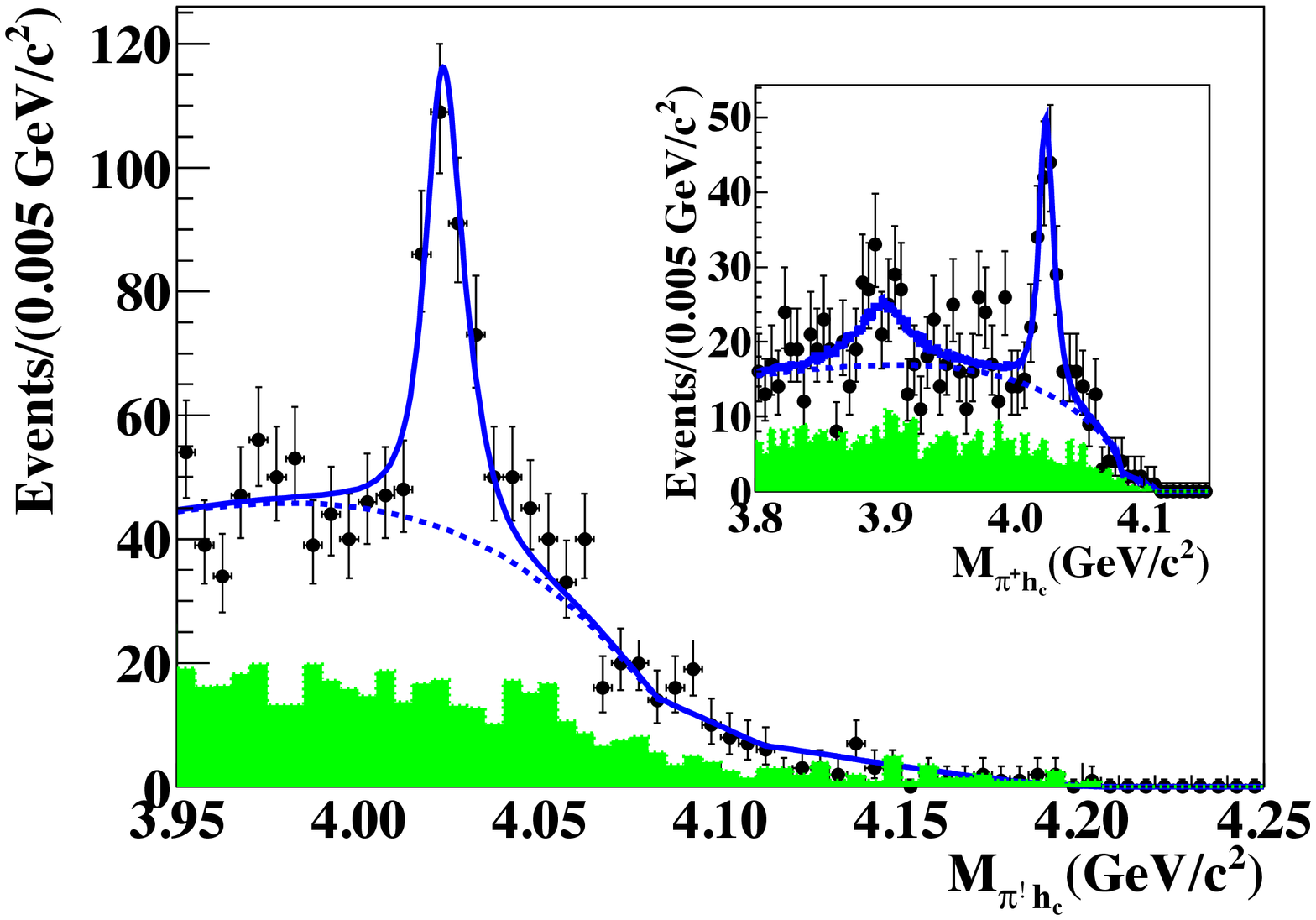}
\caption{Dalitz plot ($M^2_{\pi^+\hc}$ vs. $M^2_{\pp}$) for
selected $\EE\to \pphc$ events (left) and the $\zcp$ signal
observed in $\pi\hc$ invariant mass spectrum (right). Points with
error bars are data, the solid curves are the best fit, the shaded
histograms are the non-$\pphc$ background estimated from the
normalized $\hc$ sidebands.} \label{zcp}
\end{center}
\end{figure}

The process $\EE\to \pphc$ was observed at c.m. energies of
$3.90-4.42~\gev$~\cite{zc4020} with cross section that is similar
to that for $\EE\to \pp\jpsi$~\cite{bes3_pipijpsi_lineshape}.
Intermediate states of this three-body system were studied by
examining the Dalitz plot of the selected $\pphc$ candidate
events, similar to what was done for $\EE\to \pp\jpsi$
process~\cite{zc3900}. Although there are no clear structures in
the $\pp$ system, there is distinct evidence for an exotic
charmoniumlike structure in the $\pi^\pm\hc$ system, as clearly
evident in the Dalitz plot shown in Fig.~\ref{zcp}. This figure
also shows projections of the $M(\pi^\pm\hc)$ (two entries per
event) distribution for the signal events as well as the
background events estimated from normalized $\hc$ mass sidebands.
There is a significant peak at around $4.02~\gevcs$ (the $\zcp$),
and there are also some events at around $3.9~\gevcs$ that could
be due to the $\zc$. The mass and width of the $\zcp$ were
measured to be $(4022.9\pm 0.8\pm 2.7)~\mevcs$ and $(7.9\pm 2.7\pm
2.6)~\mev$, respectively. The statistical significance of the
$\zcp$ signal is greater than $8.9\sigma$.

In an analysis of $\EE\to \piz\piz\hc$ process, the $\zcp^0$, the
neutral isospin partner of the $\zcp^\pm$ was observed in
$\piz\hc$ system. This indicates that the $\zcp$ is an isovector
state~\cite{zc0_4020}. The open charm decay of the $\zcp$ was
observed in $\EE \to (D^*\bar{D}^*) \pi$, with a rate that is much
larger than that for its decay into
$\pi\hc$~\cite{zc4025,zc0_4025}.

\subsubsection{Nature of the $Z_c$ states} \label{Sec:z-summary}

Although many measurements have been performed on the $\zc$ and
$\zcp$, the experimental information is still not very precise.
From the experience of the $Z_c(4430)$, we know that the resonance
parameters determined from a simple one-dimensional fit to the
invariant mass distribution~\cite{Belle_zc4430} may differ from
those based on a full amplitude analysis with the interference
effects between different amplitudes considered
properly~\cite{Belle_zc4430pwa}. The same thing may happen with
the $\zc$ and $\zcp$. Amplitude analyses that are applied to the
relevant final states that extract the resonant parameters as well
as the couplings to different modes are essential to obtain more
refined information for understanding the nature of these states.
In addition, PWA also can provide measurements of Argand plots of
the $Z_c$ amplitudes, which can be used to discriminate between
different models for the $Z_c$ states.

The production of $Z_c$ states at a variety of c.m. energies can
reveal whether these states are from resonance decays or continuum
production. So far only the $\zc$ has been observed both in $\EE$
annihilation~\cite{zc3900} and in $B$-hadron
decays~\cite{D0-zc3900-1,D0-zc3900-2}. Searches for these states
in different production modes is of great importance.

These states seem to indicate that a new class of hadrons has been
observed. Since there are at least four quarks within each of
these $Z_c$ states, they have been alternatively interpreted as
compact tetraquark states, molecular states of two charmed mesons
($D\bar{D}^*+D^*\bar{D}$, $D^*\bar{D}^*$, etc.), hadro-quarkonium
states, or other multiquark configurations; in some
phenomenological studies they have been attributed to purely
kinematical effects~\cite{reviews}. Since many of these models
suffer from assumptions that are hard to prove, it is essential
that non-perturbative studies such as lattice QCD (LQCD) provide a
way to understand their underlying nature; if the $Z_c$ structures
are not purely kinematical effects, they should appear on the
lattice since they are strong interaction phenomena.

The currently available LQCD calculations that are relevant to the
$\zc$ suffer from a number of uncertainties, as has been recently
reviewed in Ref.~\cite{LiuChuan}. These include the lattice
spacing, the volume, the physical $\pi$ mass, and the channels
that are considered in the calculation.

An early lattice study performed by Prelovsek {\em et al.}
investigated the energy levels of two-meson systems including
$\pi\jpsi$, $\pi\psp$, $\rho\etac$, $D\bar{D}^*$, $D^*\bar{D}^*$,
etc., as well as tetraquark operators. However, no convincing
signals for extra new energy levels apart from the almost free
scattering states of the two mesons were
identified~\cite{Prelovsek:2014swa}. Taking $D\bar{D}^*$ as the
main relevant channel, the CLQCD collaboration performed a
calculation that was based on the single-channel L\"uscher
finite-size formalism and found a slightly repulsive interaction
between the two charmed mesons~\cite{Chen:2014afa,Chen:2015jwa}.
The results therefore do not support the possibility of a shallow
bound state for the two mesons for the pion mass values of 300,
420, and 485~MeV/$c^2$. A preliminary study using staggered quarks
finds no $J^{PC}=1^{+-}$ state distinct from the noninteracting
scattering states either, but the authors also pointed out that
future calculations with a larger interpolating operator basis may
be able to resolve this state~\cite{Lee:2014uta}.

The HALQCD collaboration studied the problem using an approach
where an effective potential is extracted from the lattice data
and then used to solve the Schr\"odinger-like
equations~\cite{Ikeda:2016zwx,Ikeda:2017mee}. A fully
coupled-channel potential for $\pi\jpsi$, $\rho\etac$, and
$D\bar{D}^*$ interactions is obtained, and a strong off-diagonal
transition between $\pi\jpsi$ and $D\bar{D}^*$ indicates that the
$\zc$ can be explained as a threshold cusp within their current
configuration ($m_\pi=400-700$~MeV/$c^2$). In order to establish a
definite conclusion on the structure of the $\zc$ in the real
world, full QCD simulations near the physical point are being
carried out~\cite{Ikeda:2016zwx,Ikeda:2017mee}.

Recently, in order to clarify the mismatch between these two
approaches, CLQCD performed a two-channel lattice study using the
two-channel Ross-Shaw effective range
expansion~\cite{Chen:2019iux}. They considered the $\pi\jpsi$ and
$D\bar{D}^*$ channels that are most strongly coupled to $\zc$ and
found that the parameters of the Ross-Shaw matrix do not seem to
support the HALQCD scenario. The parameters turn out to be large
and the Ross-Shaw $M$ matrix is far from singular, which is
required for a resonance close to the threshold. However, since
only two channels are studied, it is still not a direct comparison
with the HALQCD approach, in which three channels were studied. In
Ref.~\cite{LiuChuan}, the same three channels that the HALQCD
collaboration analyzed were considered, namely $\pi\jpsi$,
$\rho\eta_c$, and $D\bar{D}^*$. However, the final results will
not come very soon.

Whatever the nature of the $Z_c$ states turn to be, they will
teach us a lot about the hadronic structures. Unless all these
structures are purely kinematical effects (in which case it would
have to be an as yet unknown kinematic effect), they will suggest
a new category of hadrons beyond the conventional meson and baryon
picture. Recent discoveries of structures with two pairs of
charm-anticharm quarks~\cite{lhcb_x6900} and with a minimal
4-quark configuration $cs\bar{u}\bar{d}$~\cite{lhcb_x2900} confirm
this expectation. Additional searches for other conceivable states
should be performed and the theoretical consequences of these new
types of hadrons should be investigated.

\subsection{\boldmath Comprehensive study of the $\xx$ in $\EE$ collision} \label{Sec:x3872}

The $\xx$ was first observed in $B^\pm\to K^\pm \ppjpsi$ decays 17
years ago by Belle~\cite{bellex}. It was confirmed subsequently by
several other
experiments~\cite{CDFx,D0x,babarx,x3872_JPC_LHCb_2013}. Prior to
2014, the $\xx$ was only observed in $B$ meson decays and hadron
collisions. Since the quantum numbers of $\xx$ are
$J^{PC}=1^{++}$, it can be produced via radiative decays of
excited vector charmonium or charmoniumlike states such as the
$\psi$s and the $Y$s.

\begin{figure}[htbp]
\begin{center}
\includegraphics[height=5cm]{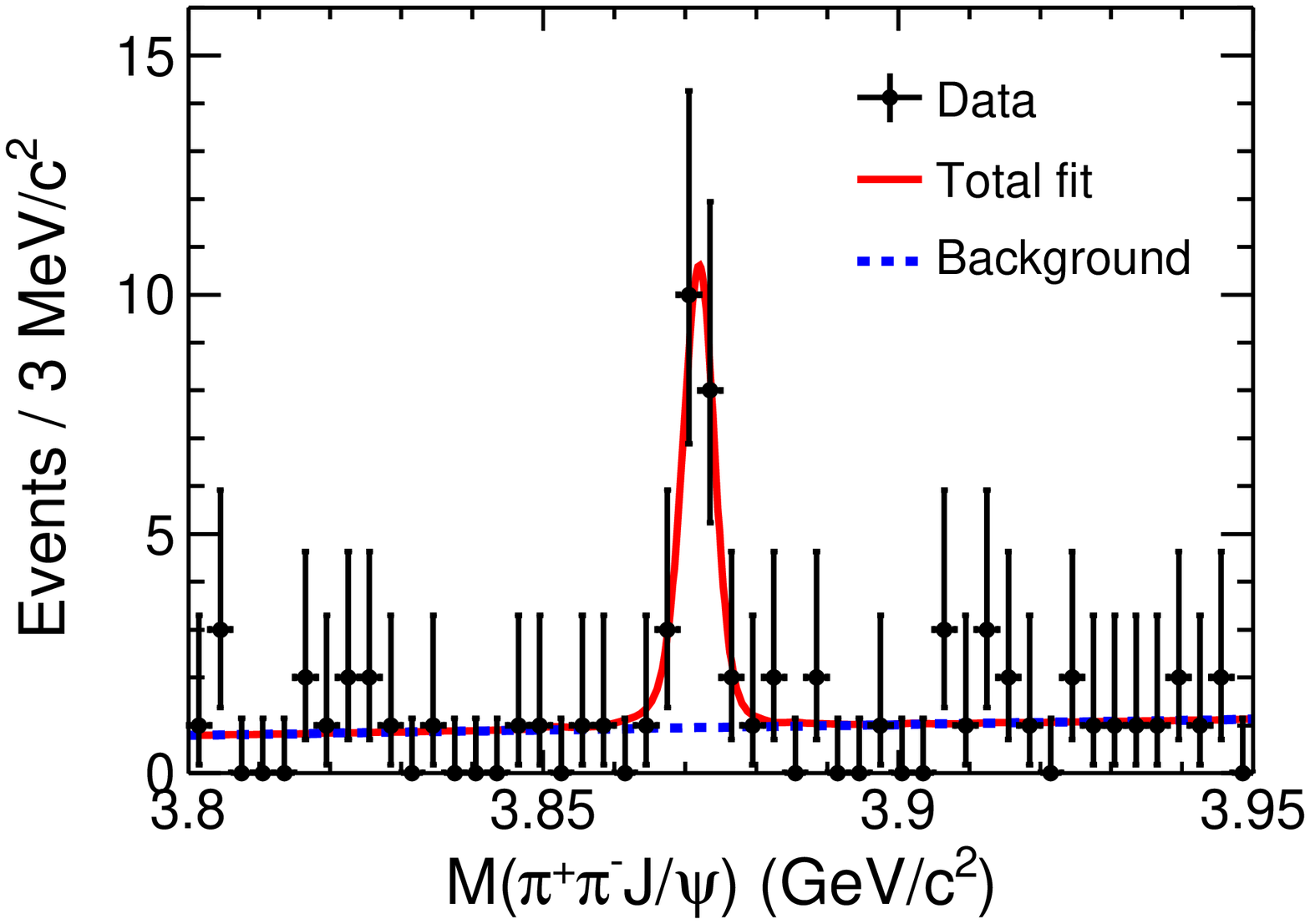}
\includegraphics[height=5cm]{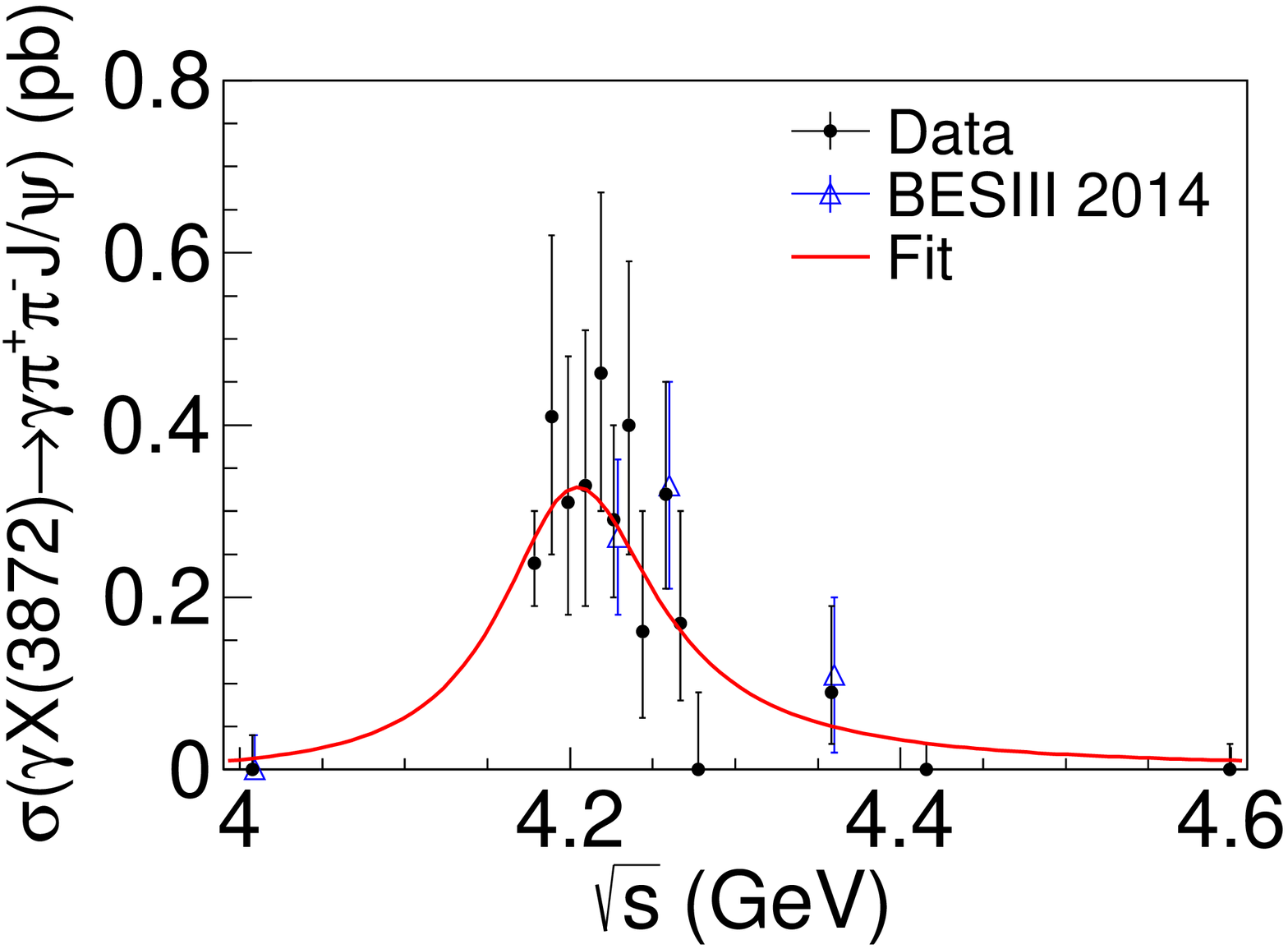}
\caption{(Left panel) Fit to the $M(\ppjpsi)$
distribution~\cite{BES3x} and (right panels) fit to
$\sigma^B[\EE\to \gamma\xx]\times \BR[\xx\to
\ppjpsi]$~\cite{BES3x_omegajpsi}. Dots/triangles with error bars
are data, and the curves are the best fits. } \label{x3872}
\end{center}
\end{figure}

The $\xx$ was observed at BESIII in the process $\EE\to \gamma\xx
\to \gamma \pp\jpsi$, $\jpsi\to \LL$~\cite{BES3x} (see the left
panel of Fig.~\ref{x3872}) and this first measurement was
subsequently improved with more data~\cite{BES3x_omegajpsi}. The
c.m. energy dependence of the product of the cross section
$\sigma[\EE\to \gamma\xx]$ and the branching fraction $\BR[\xx\to
\pp\jpsi]$ is shown in the right panel of Fig.~\ref{x3872}, where
the red curve shows the results of a fit to a BW resonance line
shape with a mass of $(4200.6^{+7.9}_{-13.3}\pm 3.0)~{\rm
MeV}/c^2$ and a width of $(115^{+38}_{-26}\pm 12)~{\rm MeV}$.
These resonance parameters are consistent with those of the
$\psi(4160)$ charmonium states~\cite{pdg} or the $Y(4220)$~(see
Sec.~\ref{Sec:y}) within errors.

Using all the data samples available at c.m. energies between
$4.0$ and $4.6~\gev$, BESIII is able to observe for the first time
significant signals of $\xx\to \omega\jpsi$~\cite{BES3x_omegajpsi}
and $\xx\to \piz\cco$~\cite{BES3x_pi0chic1}, and search for other
possible decays.

BESIII confirmed earlier observations of a large $\xx\to \ddstbn$
branching fraction and finds evidence for $\xx\to \gamma\jpsi$
with a significance of $3.5\sigma$~\cite{BES3x_anything}. No
evidence is found for the decays $\xx\to \gamma\psip$. The upper
limit on the ratio $\BR(\xx\to \gamma\psip)/ \BR(\xx\to
\gamma\jpsi) < 0.59$ is obtained at the 90\%
C.L.~\cite{BES3x_anything}, which is inconsistent with
LHCb~\cite{R_in_lhcb} and BaBar measurements~\cite{R_in_babar} but
consistent with a Belle upper limit~\cite{R_in_belle}. No
significant $\xx\to \piz\chi_{c0,2}$ signals are observed.

The hadronic transitions of the $\xx$ to low mass charmonum states
via a single pion or a rho meson violate isospin, and the large
decay rates of $\xx\to \piz\cco$ and $\rho\jpsi\to \ppjpsi$
relative to the isospin-conserved mode $\xx\to \omega\jpsi$
indicate that $\xx$ is very unlikely to be a pure charmonium
state, such as the $\chi_{c1}(2P)$. The order of magnitude larger
decay rate to $\ddstbn$ than to charmonium final state favors the
$\ddstb$ molecule interpretation of the $\xx$, as is the
relatively smaller production rate of $\xx\to \gamma\psip$
compared with $\xx\to \gamma\jpsi$, or at least that there is a
large fraction of molecular component in its wave function in
addition to a charmonium component.

BESIII measured the ratios of branching fractions for
$\xx\to\gamma\jpsi$, $\gamma\psip$, $\omega\jpsi$, $\piz\cco$,
$\ddstbn$, $\pi^0\ddb$, and $\gamma\ddb$ to that for $\xx\to
\pp\jpsi$. By combining these with the measurements of the $\xx$
properties from the $B$-factories, the authors of
Ref.~\cite{X_br_lich_ycz} obtained the absolute branching
fractions of the $\xx$ decays into six modes by globally fitting
the measurements provided by Belle, BaBar, BESIII, and LHCb
experiments (see Table~\ref{tab:xbr}). The branching fraction for
$\xx\to \pp\jpsi$ is determined to be $(4.1^{+1.9}_{-1.1})\%$,
which is in good agreement with earlier estimates in
Refs.~\cite{ycz_pic2009} and~\cite{Esposito:2014rxa}. By combining
the branching fractions of all of the observed modes, the fraction
of the unknown decays of the $\xx$ is found to be
$(31.9_{-31.5}^{+18.1})\%$, which is an important challenge for
future experimental studies of $\xx$ decays.

\begin{table*}[htbp]
\caption{\label{tab:xbr} The fitting results of the absolute
branching fractions of the $\xx$ decays~\cite{X_br_lich_ycz}. The
branching fraction of $\xx$ decays into unknown modes is
calculated from the fit results.}
\begin{tabular}{cll}
\hline\hline
Parameter index & Decay mode & Branching fraction                     \\
\hline
1 & $\xx\to \pp\jpsi$             & $(4.1^{+1.9}_{-1.1})\%$            \\
2 & $\xx\to \ddstbn$              & $(52.4^{+25.3}_{-14.3})\%$         \\
3 & $\xx\to \gamma\jpsi$          & $(1.1^{+0.6}_{-0.3})\%$            \\
4 & $\xx\to \gamma \psp$          & $(2.4^{+1.3}_{-0.8})\%$            \\
5 & $\xx\to \pi^0 \chi_{c1}$      & $(3.6^{+2.2}_{-1.6})\%$            \\
6 & $\xx\to \omega \jpsi$         & $(4.4^{+2.3}_{-1.3})\%$
\\ \hline
  & $\xx\to {\rm unknown}$        & $(31.9_{-31.5}^{+18.1})\%$         \\
\hline\hline
\end{tabular}
\end{table*}

With a very large sample of $\xx\to \ppjpsi$ events, the LHCb
experiment reported an improved measurement of its mass and a
first measurement of its width~\cite{lhcb_x_width}. Limited by its
capability of $D^{*0}$ reconstruction and mass resolution, it is
still not possible for LHCb to measure the line shape of the
resonance.

\subsection{\boldmath Commonality between the $\xx$, $\y$, and $\zc$}

With data taken with c.m. energy at and near the $\y$ resonance
peak, BESIII discovered a clear signal for $\xx$ production in
association with a $\gamma$-ray~\cite{BES3x}, as shown in
Fig.~\ref{x3872}, and a clear signal for $\zc$ production in
association with a $\pi$ meson~\cite{zc0-lineshape}, as shown in
Fig.~\ref{fig:zc0}. The c.m.-energy-dependence of the $\EE\to
\gamma\xx$ cross section is suggestive of a $\y\to \gamma\xx$
decay process, and that of $\EE\to \piz\zc^0$ cross section is
suggestive of a $\y\to \pi\zc$ decay process, these indicate that
there might be some common features to the internal structures of
the $\zc$, $\y$, and $\xx$.

Many of the models developed to interpret the nature of one of
these three states do not consider the possibility of a connection
between them. With data supplied by the BESIII experiments, some
of these models may be ruled out and others may need to be
revisited in the light of these new observations.

\section{Summary and Perspectives}
\label{Sec:Summary}

With the capability of adjusting the $\EE$ c.m. energy to the
peaks of resonances, combined with the clean experimental
environments due to near-threshold operation, BESIII is uniquely
able to perform a broad range of critical measurements of
charmonium physics, and the production and decays of many of the
nonstandard $\xyz$ states, as discussed above in the context of
the studies of the $\xx$, $Y(4220)$, $\zc$ and $\zcp$.
Table~\ref{exps_comp} shows the operating times associated the
discoveries of the $\xyz$ states at BESIII and other experiments,
including the previous generation $B$-factories BaBar and Belle,
and the new generation super-$B$-factories LHCb and Belle II.
BESIII's special advantages for studying the $\xyz$ states are
evident.

\begin{table}
  \centering
\caption{The numbers of observed events of discovery modes of the
$\xyz$ states at the BESIII and other experiments. Here the states
are detected with $\xx\to \pp\jpsi$, $\y\to \pp\jpsi$, $\zc^\pm\to
\pi^\pm\jpsi$, $\zcp\to \pi^\pm h_c$, and $Y(4660)\to \pp\psp$.
The numbers for Belle II experiments are a simple scale according
to those of Belle experiment. ``--" means no measurement
available. BESIII can detect other decay modes of these states
while other experiments can barely do. }\label{exps_comp}
\begin{tabular}{llccccc}
  \hline\hline
 Experiment & Data taking time& $\xx$       & $\y$ & $\zc$ & $\zcp$ & $Y(4660)$ \\\hline
  BESIII   & 3 months         &  20         & 6,000 & 1,300 & 180 &  250 \\\hline
  BaBar    & 1999-2008        &  90         &  270 &   80 &  -- &  45 \\
  Belle    & 1999-2010        & 170         &  550 &  160 &  -- &  90 \\\hline
  LHCb     & 2011-12~($B$ decays)     & 4,000   &  -- &   -- &  -- & -- \\
           & 2011-18~($pp$ collision) & 16,000  &  -- &   -- &  -- & -- \\
  Belle II & 2019-30          &          8,000 & 28,000  &  8,000  &  -- & 5,000 \\
  \hline\hline
\end{tabular}
\end{table}

We emphasize here that BESIII measured all the known decay modes
of the $\xx$ and discovered its new decay modes even though the
numbers of produced $\xx$ events are much smaller than those of
other experiments. This is because the very clean experimental
environment of $\EE$ collisions in the $\tau$-charm threshold
energy region uniquely facilitates the isolation of signals for
$\xx$ decays into final states with one or more photons with high
efficiency. This is especially true for final states that contain
an $\hc$ charmonium state like the BESIII discovery of the $\zcp$
state and measurements of $Y(4220)$ and $Y(4390)\to \pphc$ decays.
Neither the BaBar and Belle $B$-factory nor the LHCb experiment
has ever seen an $\hc$ signal.

BESIII has produced a considerable amount of information about the
$\xyz$ and the conventional charmonium states. In addition, there
are still data that are still being analyzed and more data that
will be accumulated at other c.m. energies~\cite{fop,bes3_white}.
Analyses with these additional data samples will provide an
improved understanding of the $\xyz$ states, especially the $\xx$,
$\y$, $\zc$, and $\zcp$. The maximum c.m. energy accessible at
BEPCII was upgraded from 4.6 to 4.9~GeV in 2019, and a
3.5~fb$^{-1}$ sample of data between 4.6 and 4.7~GeV was
accumulated in the 2019-20 running period, with more data planned
for the future. This will enable a full coverage of the
$Y(4660)$~\cite{belle_y4660} resonance and a search for possible
higher mass vector mesons and states with other quantum numbers,
as well as improved measurements of their properties.

At the same time, other experiments will also supply information
on these states. At the LHCb, in addition to the 3~fb$^{-1}$ data
at 7 and 8~TeV that have been used for most of their published
analyses, there is a 6~fb$^{-1}$ data sample at 13~TeV that is
being used for improved analyses of many of the topics discussed
above such as the $\xx$ decay properties and the searches for the
$Y$ and $\zc$ states in $B$ decays.

Belle II~\cite{belle2} has collected about 70~fb$^{-1}$ data by
mid-2020, and will accumulate 50~ab$^{-1}$ data at the
$\Upsilon(4S)$ peak by the end of 2030. These data samples can be
used to study the $\xyz$ and charmonium states in many different
ways~\cite{PBFB}, among which ISR can produce events in the same
energy range covered by BESIII. A 50~ab$^{-1}$ Belle II data
sample will correspond to 2.0--2.8~fb$^{-1}$ data for every 10~MeV
from 4--5~GeV. Similar statistics will be available for modes like
$\EE\to \ppjpsi$ at Belle II and BESIII (after considering the
fact that Belle II has lower efficiency). Belle II has the
advantage that data at different energies will be accumulated at
the same time, making the analysis much simpler than at BESIII.

There are two super $\tau$-charm factories proposed, the STC in
China~\cite{HIEPA} and the SCT in Russia~\cite{SCT_charm2018}.
Both machines would run at c.m. energies of up to 5~GeV or higher
with a peak luminosity of $10^{35}$~cm$^{-2}$s$^{-1}$ which is a
factor of 100 improvement over the BEPCII. This would enable
systematic studies of the $\xyz$ and charmonium states with
unprecedented precision.

\section*{Acknowledgments}

I thank my BESIII collaborators for producing these fantastic
results presented in this review, I thank Steve Olsen for comments
and suggestions on the manuscript. This work is supported in part
by National Key Research and Development Program of China under
Contract No.~2020YFA0406300, National Natural Science Foundation
of China (NSFC) under contract Nos. 11961141012, 11835012, and
11521505; and the CAS Center for Excellence in Particle Physics
(CCEPP).

\bigskip
\noindent {\bf\it Note added:} After submission of this
manuscript, BESIII announced observation of a near-threshold
structure in the $K^+$ recoil-mass spectra in $\EE\to
K^+(D^-_sD^{*0}+D^{*-}_sD^{0})$ which could be the strange partner
of the $\zc$ with the $d$ quark replaced with an $s$
quark~\cite{zcs}.


\begin{thebibliography}{000}

\bibitem{ting}
  J.~J.~Aubert {\it et al.} [E598 Collaboration],
  Phys.\ Rev.\ Lett.\  {\bf 33}, 1404 (1974).

\bibitem{richter}
  J.~E.~Augustin {\it et al.} [SLAC-SP-017 Collaboration],
  Phys.\ Rev.\ Lett.\  {\bf 33}, 1406 (1974).

\bibitem{pdg}
 P.~A.~Zyla {\it et al.} (Particle Data Group), Prog. Theor. Exp. Phys.
 \textbf{2020}, 083C01 (2020).

\bibitem{eichten} E.~Eichten, K.~Gottfried, T.~Kinoshita, K.~D.~Lane and T.~M.~Yan,
  Phys.\ Rev.\ D {\bf 17}, 3090 (1978).

\bibitem{godfrey}   S.~Godfrey and N.~Isgur,
  Phys.\ Rev.\ D {\bf 32}, 189 (1985).

\bibitem{barnes}  T.~Barnes, S.~Godfrey and E.~S.~Swanson,
  Phys.\ Rev.\ D {\bf 72}, 054026 (2005).

\bibitem{Jaffe:2004ph}
  R.~L.~Jaffe,
  Phys.\ Rept.\  {\bf 409}, 1 (2005).

\bibitem{klempt}
 For a review, see E.~Klempt and A.~Zaitsev,
 Phys.\ Rept.\  {\bf 454}, 1 (2007).

\bibitem{babar}
 B.~Aubert {\it et al.} [BaBar Collaboration],
 Nucl.\ Instrum.\ Methods Phys.\ Res.\, Sect.\ A {\bf 479}, 1 (2002).

\bibitem{belle}
 A.~Abashian {\it et al.},
 Nucl.\ Instrum.\ Methods Phys.\ Res.\, Sect.\ A {\bf 479}, 117 (2002).

\bibitem{PBFB}
 A.~J.~Bevan {\it et al.} [BaBar and Belle Collaborations],
 Eur.\ Phys.\ J.\ C {\bf 74}, 3026 (2014).

\bibitem{reviews}
  For recent reviews, see
  N.~Brambilla, S.~Eidelman, C.~Hanhart, A.~Nefediev, C.~P.~Shen, C.~E.~Thomas, A.~Vairo and C.~Z.~Yuan,
  Phys. Rept. \textbf{873}, 1 (2020);
  F.~K.~Guo, C.~Hanhart, U.~G.~Mei{\ss}ner, Q.~Wang, Q.~Zhao and B.~S.~Zou,
  Rev.\ Mod.\ Phys.\  {\bf 90}, 015004 (2018);
  H.~X.~Chen, W.~Chen, X.~Liu and S.~L.~Zhu,
  Phys.\ Rept.\  {\bf 639}, 1 (2016);
  N.~Brambilla {\em et al.},
  Eur.\ Phys.\ J.\ C {\bf 71}, 1534 (2011).

\bibitem{bes3}
M. Ablikim {\em et al.} [BESIII Collaboration],
Nucl.\ Instrum.\ Methods Phys.\ Res.\, Sect.\ A {\bf 614}, 345 (2010).

\bibitem{lhcb_detector}
A.~A.~Alves, Jr. {\it et al.} [LHCb Collaboration],
JINST {\bf 3}, S08005 (2008);
R.~Aaij {\it et al.} [LHCb Collaboration],
Int.\ J.\ Mod.\ Phys.\ A {\bf 30}, 1530022 (2015).

\bibitem{BESIII_YB}
D.~M.~Asner {\it et al.},
Int.\ J.\ Mod.\ Phys.\ A {\bf 24}, S1 (2009).

\bibitem{zc3900}
M.~Ablikim {\it et al.} [BESIII Collaboration],
Phys.\ Rev.\ Lett.\  {\bf 110}, 252001 (2013).

\bibitem{lum_songwm}
M.~Ablikim {\it et al.} [BESIII Collaboration],
Chin.\ Phys.\ C {\bf 39}, 093001 (2015).

\bibitem{lum_Rscan}
M.~Ablikim {\it et al.} [BESIII Collaboration],
Chin.\ Phys.\ C {\bf 41}, 063001 (2017).

\bibitem{3d2-mass}
  W.~Kwong, J.~L.~Rosner and C.~Quigg,
  Ann.\ Rev.\ Nucl.\ Part.\ Sci.\  {\bf 37}, 325 (1987);
  D.~Ebert, R.~N.~Faustov and V.~O.~Galkin,
  Phys.\ Rev.\ D {\bf 67}, 014027 (2003);
  B.~Q.~Li and K.~T.~Chao,
  Phys.\ Rev.\ D {\bf 79}, 094004 (2009);
  M.~Blank and A.~Krassnigg,
  Phys.\ Rev.\ D {\bf 84}, 096014 (2011).

\bibitem{ratio}   C.~F.~Qiao, F.~Yuan and K.~T.~Chao,
  Phys.\ Rev.\ D {\bf 55}, 4001 (1997).

\bibitem{BES3x3823}
M.~Ablikim {\it et al.} [BESIII Collaboration],
Phys.\ Rev.\ Lett.\  {\bf 115}, 011803 (2015).

\bibitem{lhcb-3d3}
R.~Aaij \textit{et al.} [LHCb Collaboration],
JHEP \textbf{07}, 035 (2019).

\bibitem{bes3_pipihc_lineshape}
M.~Ablikim {\it et al.} [BESIII Collaboration],
Phys.\ Rev.\ Lett.\  {\bf 118}, 092002 (2017).

\bibitem{bes3_etacbr}
M.~Ablikim \textit{et al.} [BESIII Collaboration],
Phys. Rev. D \textbf{100}, 012003 (2019).

\bibitem{bes3_etacp}
M.~Ablikim \textit{et al.} [BES Collaboration],
Phys. Rev. Lett. \textbf{109}, 042003 (2012).

\bibitem{gaoky} See the compilation of the
results in K.~Gao, ``Study of Radiative Decays of $\psi(2S)$
Mesons'', arXiv:0909.2812 [hep-ex].

\bibitem{zhaoq}
  G.~Li, Q.~Zhao,
  Phys.\ Lett.\  B {\bf 670}, 55 (2008).

\bibitem{mabq} T.~Peng and B.~Ma,
Eur. Phys. J. A \textbf{48}, 66 (2012).

\bibitem{cbal} C.~Edwards {\it et al.} (Crystal Ball Collaboration),
Phys.\ Rev.\ Lett.\  {\bf 48}, 70 (1982).

\bibitem{ycz} C.~Z.~Yuan, ``Search for $\etacp$ and study of
$\chi_{cJ}$ decays using $\psp$ data'', Ph.D thesis, Institute of
High Energy Physics, Chinese Academy of Sciences, 1997.

\bibitem{initial_psp}
C.~Z.~Yuan,
Mod. Phys. Lett. A \textbf{35}, 2030009 (2020).

\bibitem{etacp_cleo_c} D.~Cronin-Hennessy {\it et al.}
(CLEO Collaboration), Phys. Rev. D {\bf 81}, 052002 (2010).

\bibitem{cbal_etac}
R.~Partridge {\em et al.} [Crystal Ball Collaboration],
Phys. Rev. Lett. \textbf{45}, 1150 (1980).

\bibitem{belle-3d2}   V.~Bhardwaj {\it et al.} [Belle Collaboration],
  Phys.\ Rev.\ Lett.\  {\bf 111}, 032001 (2013).

\bibitem{bellex} S. K. Choi {\em et al.} [Belle Collaboration],
\Journal\PRL{91}{262001}{2003}.

\bibitem{babar_y4260}
B.~Aubert {\it et al.} [BaBar Collaboration],
Phys.\ Rev.\ Lett.\  {\bf 95}, 142001 (2005).

\bibitem{babar_y4360}
B.~Aubert {\it et al.} [BaBar Collaboration],
Phys.\ Rev.\ Lett.\ {\bf 98}, 212001 (2007).

\bibitem{belle_y4660}
X.~L.~Wang {\it et al.} [Belle Collaboration],
Phys.\ Rev.\ Lett.\ {\bf 99}, 142002 (2007).

\bibitem{bes2_psis}
  M.~Ablikim {\it et al.} [BES Collaboration],
  Phys.\ Lett.\ B {\bf 660}, 315 (2008).

\bibitem{uglov-kmatrix}
T.~V.~Uglov, Y.~S.~Kalashnikova, A.~V.~Nefediev, G.~V.~Pakhlova
and P.~N.~Pakhlov,
JETP Lett.\  {\bf 105}, 1 (2017).

\bibitem{belley}
C.~Z.~Yuan {\it et al.} [Belle Collaboration],
Phys.\ Rev.\ Lett.\  {\bf 99}, 182004 (2007).

\bibitem{babar_y4260_new}
J.~P.~Lees {\em et al.} [BaBar Collaboration],
Phys.\ Rev.\ D {\bf 86}, 051102(R) (2012).

\bibitem{belley_new}
Z.~Q.~Liu {\it et al.} [Belle Collaboration],
Phys.\ Rev.\ Lett.\  {\bf 110}, 252002 (2013).

\bibitem{cleo_isry}
Q.~He \textit{et al.} [CLEO Collaboration],
Phys. Rev. D \textbf{74}, 091104 (2006).

\bibitem{pdg2016}
C.~Patrignani {\it et al.} [Particle Data Group], Chin.\ Phys.\ C
{\bf 40}, 100001 (2016).

\bibitem{bes3_pipijpsi_lineshape}
M.~Ablikim {\it et al.} [BESIII Collaboration],
Phys.\ Rev.\ Lett.\  {\bf 118}, 092001 (2017).

\bibitem{bes3_omegachic0}
M.~Ablikim {\it et al.} [BESIII Collaboration],
Phys.\ Rev.\ Lett.\  {\bf 114}, 092003 (2015).

\bibitem{bes3_omegachic0_new}
M.~Ablikim \textit{et al.} [BESIII Collaboration],
Phys. Rev. D \textbf{99}, 091103 (2019).

\bibitem{bes3_ddstarpi}
M.~Ablikim \textit{et al.} [BESIII Collaboration],
Phys. Rev. Lett. \textbf{122}, 102002 (2019).

\bibitem{bes3_white}
M.~Ablikim, \textit{et al.},
Chin. Phys. C \textbf{44}, 040001 (2020).

\bibitem{gaoxy}
X.~Y.~Gao, C.~P.~Shen and C.~Z.~Yuan,
Phys.\ Rev.\ D {\bf 95}, 92007 (2017).

\bibitem{zhenghq}
  Q.~F.~Cao, H.~R.~Qi, G.~Y.~Tang, Y.~F.~Xue and H.~Q.~Zheng,
  arXiv:2002.05641 [hep-ph].

\bibitem{chenying}
Y.~Chen, W.~F.~Chiu, M.~Gong, L.~C.~Gui and Z.~ Liu,
Chin.\ Phys.\ C {\bf 40}, 081002 (2016).


\bibitem{Belle_zc4430}
  S.~K.~Choi {\it et al.}  [Belle Collaboration],
  Phys.\ Rev.\ Lett.\  {\bf 100}, 142001 (2008).

\bibitem{Belle_zc4430pwa}
  K.~Chilikin {\it et al.}  [Belle Collaboration],
  Phys.\ Rev.\ D {\bf 88}, 074026 (2013).

\bibitem{LHCb_zc4430}     R.~Aaij {\it et al.} [LHCb Collaboration],
  Phys.\ Rev.\ Lett.\  {\bf 112}, 222002 (2014).

\bibitem{zc4020}
M.~Ablikim {\it et al.} [BESIII Collaboration],
Phys.\ Rev.\ Lett.\  {\bf 111}, 242001 (2013).

\bibitem{seth_zc} T.~Xiao, S.~Dobbs, A.~Tomaradze and K.~K.~Seth,
  Phys.\ Lett.\ B {\bf 727}, 366 (2013).

\bibitem{zc3900_jpc}
  M.~Ablikim {\it et al.} [BESIII Collaboration],
  Phys.\ Rev.\ Lett.\  {\bf 119}, 072001 (2017).

\bibitem{zc3885st}
  M.~Ablikim {\it et al.}  [BESIII Collaboration],
  Phys.\ Rev.\ Lett.\  {\bf 112}, 022001 (2014).

\bibitem{zc3885dt}
  M.~Ablikim {\it et al.} [BESIII Collaboration],
  Phys.\ Rev.\ D {\bf 92}, 092006 (2015).

\bibitem{Ablikim:2015tbp}
M.~Ablikim \textit{et al.} [BESIII Collaboration],
Phys. Rev. Lett. \textbf{115}, 112003 (2015).

\bibitem{Ablikim:2015gda}
M.~Ablikim \textit{et al.} [BESIII Collaboration],
Phys. Rev. Lett. \textbf{115}, 222002 (2015).

\bibitem{zc0-lineshape}
M.~Ablikim \textit{et al.} [BESIII Collaboration],
Phys. Rev. D \textbf{102}, 012009 (2020).

\bibitem{Zc_jpsieta_bes3}
  M.~Ablikim {\it et al.} [BESIII Collaboration],
  Phys.\ Rev.\ D {\bf 92}, 012008 (2015).

\bibitem{Zc_omegapi_bes3}
  M.~Ablikim {\it et al.} [BESIII Collaboration],
  Phys.\ Rev.\ D {\bf 92}, 032009 (2015).

\bibitem{Zc_kkpi_bes3}
M.~Ablikim \textit{et al.} [BESIII Collaboration],
Phys. Rev. D \textbf{99}, 012003 (2019).

\bibitem{zc0_4020}
M.~Ablikim {\it et al.} [BESIII Collaboration],
Phys.\ Rev.\ Lett.\  {\bf 113}, 212002 (2014).

\bibitem{zc4025}   M.~Ablikim {\it et al.}  [BESIII Collaboration],
  Phys.\ Rev.\ Lett.\  {\bf 112}, 132001 (2014).

\bibitem{zc0_4025}   M.~Ablikim {\it et al.} [BESIII Collaboration],
  Phys.\ Rev.\ Lett.\  {\bf 115}, 182002 (2015).

\bibitem{D0-zc3900-1}
V.~M.~Abazov \textit{et al.} [D0 Collaboration],
Phys. Rev. D \textbf{98}, 052010 (2018).

\bibitem{D0-zc3900-2}
V.~M.~Abazov \textit{et al.} [D0 Collaboration],
Phys. Rev. D \textbf{100}, 012005 (2019).

\bibitem{LiuChuan}
C.~Liu, L.~Liu and K.~Zhang,
Phys.\ Rev.\ D \textbf{101}, 054502 (2020).

\bibitem{Prelovsek:2014swa}
S.~Prelovsek, C.~Lang, L.~Leskovec and D.~Mohler,
Phys. Rev. D \textbf{91}, 014504 (2015).

\bibitem{Chen:2014afa}
Y.~Chen, M.~Gong, Y.~H.~Lei, N.~Li, J.~Liang, C.~Liu, H.~Liu,
J.~L.~Liu, L.~Liu, Y.~F.~Liu, Y.~B.~Liu, Z.~Liu, J.~P.~Ma,
Z.~L.~Wang, Y.~B.~Yang and J.~B.~Zhang,
Phys. Rev. D \textbf{89}, 094506 (2014).

\bibitem{Chen:2015jwa}
Y.~Chen \textit{et al.} [CLQCD],
Phys. Rev. D \textbf{92}, 054507 (2015).

\bibitem{Lee:2014uta}
S.~h.~Lee \textit{et al.} [Fermilab Lattice and MILC],
[arXiv:1411.1389 [hep-lat]].

\bibitem{Ikeda:2016zwx}
Y.~Ikeda \textit{et al.} [HALQCD],
Phys. Rev. Lett. \textbf{117}, 242001 (2016).

\bibitem{Ikeda:2017mee}
Y.~Ikeda [HALQCD],
J. Phys. G \textbf{45}, 024002 (2018).

\bibitem{Chen:2019iux}
T.~Chen \textit{et al.} [CLQCD],
Chin. Phys. C \textbf{43}, 103103 (2019).

\bibitem{lhcb_x6900}
R.~Aaij \textit{et al.} [LHCb Collaboration],
[arXiv:2006.16957 [hep-ex]].

\bibitem{lhcb_x2900} R.~Aaij \textit{et al.} [LHCb Collaboration],
[arXiv:2009.00026 [hep-ex]].

\bibitem{CDFx} D. Acosta {\em et al.} [CDF Collaboration],
\Journal\PRL{93}{072001}{2004}.

\bibitem{D0x} V. M. Abazov {\em et al.} [D0 Collaboration],
\Journal\PRL{93}{162002}{2004}.

\bibitem{babarx} B. Aubert {\em et al.} [BaBar Collaboration],
\Journal\PRD{71}{071103}{2005}.

\bibitem{x3872_JPC_LHCb_2013}
  R.~Aaij {\it et al.} [LHCb Collaboration],
  Phys.\ Rev.\ Lett.\  {\bf 110}, 222001 (2013).

\bibitem{BES3x}
M.~Ablikim {\it et al.} [BESIII Collaboration],
Phys.\ Rev.\ Lett.\  {\bf 112}, 092001 (2014).

\bibitem{BES3x_omegajpsi}
M.~Ablikim \textit{et al.} [BESIII Collaboration],
Phys. Rev. Lett. \textbf{122}, 232002 (2019).

\bibitem{BES3x_pi0chic1}
M.~Ablikim \textit{et al.} [BESIII Collaboration],
Phys. Rev. Lett. \textbf{122}, 202001 (2019).

\bibitem{BES3x_anything}
M.~Ablikim \textit{et al.} [BESIII Collaboration],
Phys. Rev. Lett. \textbf{124}, 242001 (2019).

\bibitem{R_in_lhcb}
  R.~Aaij {\it et al.} [LHCb Collaboration],
  Nucl.\ Phys.\ B {\bf 886}, 665 (2014).

\bibitem{R_in_babar}
B.~Aubert \textit{et al.} [BaBar Collaboration],
Phys. Rev. Lett. \textbf{102}, 132001 (2009).

\bibitem{R_in_belle}
V.~Bhardwaj \textit{et al.} [Belle Collaboration],
Phys. Rev. Lett. \textbf{107}, 091803 (2011).

\bibitem{X_br_lich_ycz}
C.~Li and C.~Z.~Yuan,
Phys. Rev. D \textbf{100}, 094003 (2019).

\bibitem{ycz_pic2009}
  C.~Z.~Yuan [for the Belle Collaboration], ``Exotic Hadrons,''
  talk at the 29th International Conference on Physics in Collision (PIC 2009),
  Kobe, Japan, August 30-September 2,
  arXiv:0910.3138 [hep-ex].

\bibitem{Esposito:2014rxa}
  A.~Esposito, A.~L.~Guerrieri, F.~Piccinini, A.~Pilloni and A.~D.~Polosa,
  Int.\ J.\ Mod.\ Phys.\ A {\bf 30}, 1530002 (2015).

\bibitem{lhcb_x_width}
R.~Aaij \textit{et al.} [LHCb Collaboration],
[arXiv:2005.13419 [hep-ex]].

\bibitem{fop}  Chang-Zheng Yuan,
 Front.\ Phys.\ {\bf 10}, 101401 (2015).

\bibitem{belle2}
  T.~Abe {\it et al.}  [Belle II Collaboration],
  arXiv:1011.0352 [physics.ins-det].

\bibitem{HIEPA} Z.~G.~Zhao, talk at the
``International Workshop on Physics at Future High Intensity
Collider \@ 2-7~GeV in China", January 13-16, 2015, University of
Chinese Academy of Sciences (UCAS), Hefei, China.

\bibitem{SCT_charm2018} Eugeny Levichev, talk at
``The 9th International Workshop on Charm Physics", May 21 to 25,
2018, Novosibirsk, Russia.
\url{https://indico.inp.nsk.su/event/10/session/1/contribution/65/material/slides/0.pdf}

\bibitem{zcs}
M.~Ablikim \textit{et al.} [BESIII Collaboration],
[arXiv:2011.07855 [hep-ex]].

\end{thebibliography}
\end{document}